%% file: main.tex
\documentclass{aa}  

\usepackage{graphicx}
\usepackage{color}
\usepackage{txfonts}
\usepackage{hyperref}
\usepackage[para,online,flushleft]{threeparttable}
\usepackage{flushend}
\usepackage{deluxetable}
\usepackage[rgb]{xcolor}
\usepackage{pdfcomment}
\begin{document}

   \title{Compact stellar systems in the polar ring galaxies NGC\,4650A and NGC\,3808B: Clues to polar disk formation}

%   \subtitle{I. Overviewing the $\kappa$-mechanism}

   \author{Yasna Ordenes-Brice\~no
          \inst{1},
          Iskren Y. Georgiev\inst{2}
		  \and
Thomas H. Puzia\inst{1}
		  \and
          Paul Goudfrooij\inst{3}
          \and
          Magda Arnaboldi\inst{4}
          }

   \institute{Institute of Astrophysics, Pontificia Universidad Cat\'olica de Chile, Av. Vicu\~na Mackenna 4860, 7820436 Santiago, Chile\\
              \email{yordenes@astro.puc.cl}
             \thanks{Additional e-mail contact: georgiev@mpia.de}
         \and
Max Plank Institute for Astronomy, K\"onigstuhl 17, 69117 Heidelberg, Germany
          \and
          Space Telescope Science Institute, 3700 San Martin Drive, Baltimore MD-21218, Maryland, USA
          \and
          European Southern Observatory, Karl-Schwarzschild-Strasse 2, D-85748 Garching bei München
}

   \date{Received 21 July 2015 / Accepted 10 November 2015}

% \abstract{}{}{}{}{} 
% 5 {} token are mandatory
 
  \abstract
   {Polar ring galaxies (PRGs) are composed of two kinematically distinct and nearly orthogonal components, a host galaxy (HG) and a polar ring/disk (PR). The HG usually contains an older stellar population than the PR. The suggested formation channel of PRGs is still poorly constrained. Suggested options are merger, gas accretion, tidal interaction, or a combination of both.
   }
   {To constrain the formation scenario of PRGs, we study the compact stellar systems (CSSs) in two PRGs at different evolutionary stages: NGC\,4650A with well-defined PR, and NGC\,3808\,B, which is in the process of PR formation.
   }
   {We use archival HST/WFPC2 imaging in the $F450W,F555W$, or $F606W$ and $F814W$ filters. Extensive completeness tests, PSF-fitting techniques, and color selection criteria are used to select cluster candidates. Photometric analysis of the CSSs was performed to determine their ages and masses using stellar population models at a fixed metallicity.
   }
   {Both PRGs contain young CSSs ($<\!1$\,Gyr) with masses of up to 5\,$\times\,10^6M_\odot$, mostly located in the PR and along the tidal debris. The most massive CSSs may be progenitors of metal-rich globular clusters or ultra compact dwarf (UCD) galaxies. We identify one such young UCD candidate, NGC\,3808\,B-8, and measure its size of $r_{\rm eff}=25.23^{+1.43}_{-2.01}$\,pc. We reconstruct the star formation history of the two PRGs and find strong peaks in the star formation rate (SFR, $\simeq$200M$_\odot$/yr) in NGC\,3808\,B, while NGC\,4650\,A shows milder (declining) star formation (SFR\,$<10M_\odot$\,/yr). This difference may support different evolutionary paths between these PRGs.
  } 
{The spatial distribution, masses, and peak star formation epoch of the clusters in NGC\,3808 suggest for a tidally triggered star formation. Incompleteness at old ages prevents us from probing the SFR at earlier epochs of NGC\,4650\,A, where we observe the fading tail of CSS formation. This also impedes us from testing the formation scenarios of this PRG. 
}

\titlerunning{Compact stellar systems in two PRGs: Clues to polar disk formation}
\authorrunning{Ordenes-Brice\~no et al.}
   \maketitle
%
%________________________________________________________________

\section{Introduction}\label{Sect:Intro}
Understanding the physical processes driving galaxy formation and evolution over a Hubble time is a key question in astrophysics. 
The role of galaxy merging and interaction is one of the main mechanisms under consideration that shape the evolutionary history of galaxies.
Their main manifestation is in the observed increased galaxy star formation rate (SFR) accompanied by the formation of a significant amount of new stars and star clusters (SC); \cite[e.g.;][]{Schweizer83,Schweizer87,Whitmore99,Goudfrooij07,Bournaud10,Duc13}. Massive star clusters form in a very short timescale and all their stars share virtually the same age and chemical composition. Therefore, individually and as a system, the SCs' properties offer one of the best tools to reconstruct the star formation history of their host Galaxy, which cannot be resolved into individual stars \cite[e.g.;][]{Goudfrooij01,Puzia02,Hempel07,Bastian08,Fedotov11,Georgiev12}.

One of the most interesting morphological types of galaxies, which bear the signs of intense interactions, are the polar ring galaxies \cite[PRGs;][]{Whitmore90}. PRGs have a peculiar appearance characterized by a ring/disk orbiting the central galaxy in a nearly orthogonal plane. The host galaxy is typically a spheroid, but there are cases where it is also a spiral disk \cite[e.g., NGC 660;][]{Karataeva04}. 
In some cases, the polar ring is very extended, two to three times the diameter of the central galaxy and it is rotating, in which case it is called a polar disk \cite[e.g, NGC\,4650A;][]{Iodice06}.

Several formation scenarios are proposed for PRGs. The three main scenarios are: ($i$) The merger scenario, where at least one of the galaxies is gas rich and the collision is in a perpendicular orbit \citep{Bekki98,Bournaud03}; ($ii$) the tidal accretion scenario, where mass is accreted from a nearby companion \citep{Reshetnikov97,Bournaud03}; and ($iii$) cold gas accretion from the intergalactic medium \citep{Maccio06,Brook08}.  Simulations have demonstrated that these scenarios are able to form stable polar ring structures. However, each individual channel leads to different and often distinct observable properties, which can be empirically tested.  The aim of this study is to provide constraints on models of PRG formation by studying two of these systems. In particular, the tidal interaction scenario has received observational support suggesting that it is an effective mechanism to form PRGs, where one of the best examples for PRGs in the making are NGC\,3808A/B and NGC\,6285/86, \citep{Reshetnikov96}.
Motivated by the fact that galaxy interactions are often accompanied by the formation of massive star clusters, we study the star cluster populations of two PRG systems in apparently different evolutionary stages: NGC\,4650A and NGC\,3808B. We find a well-defined polar ring in the former and one in the making in the latter (see \S\,\ref{Sect:Galaxy sample} Fig. \ref{fig:selection}).

The integrated-light properties of star clusters can be used to reconstruct the galaxy star formation history (SFH). A direct method for deciphering the galaxy SFH is using the relation between the galaxy SFR and the most massive cluster at a given epoch \citep{Larsen02,Weidner04,Maschberger07,Bastian08}. This method has been successfully tested by \cite{Maschberger11} in deriving the SFH of the Large Magellanic Cloud from the most massive cluster at a given age, which yields very comparable SFH as derived from resolved CMD analysis \citep{Harris09}. 
This approach has received recent attention in reconstructing the SFHs of massive, post-interacting galaxies from their most massive star clusters at a given age \citep{Georgiev12}. A number of studies have shown that, in general, galaxies with higher SFR (or surface SFR, $\Sigma_{\rm SFR}$) form more luminous (massive) and, on average, more star clusters \citep{Meurer95,Zepf99,Larsen&Richtler00,Bastian08,Adamo11,Cook12}, however, these results might be affected by small-number statistics at the high-mass end of the mass distribution \citep{Whitmore03,Larsen04,Bastian08}. See also a review and references in \cite{Bastian13}.
Here we use the most massive cluster at a given age as a proxy for the epoch of peak SFR in our sample PRGs. This allows us to gain insights into their evolutionary path and star formation history. Specifically, we use the peak star formation epochs as a look back time clock to estimate the age of the event that led to the formation of the outer structure, i.e., the polar ring. That is, we use the time of peak SFR and projected distances from the nearest massive galaxy to calculate tangential and orbital velocities and see whether they are consistent with $i)$ a fly-by tidal interaction or $ii)$ bound orbital motions.

In \S\,\ref{Sect:Data} we present the sample PRGs and the reduction of archival Hubble Space Telescope (HST) imaging data with the Wide Field and Planetary Camera 2 (WFPC2), the star clusters' selection, photometry, and completeness tests. In \S\,\ref{Sect:Analysis} we analyse, the color-color and color-magnitude diagrams of the cluster candidates. We discuss the SFHs of the two PRGs in \S\,\ref{Sect:PRG SFHs} and what implications our results have on their formation in \S\,\ref{Sect: Discussion}.~Finally~we~summarize~our findings in~\S\,\ref{Sect:Conclusions}.

\section{Data sample, image reduction, and photometry}\label{Sect:Data}

\subsection{The sample galaxies}\label{Sect:Galaxy sample}

The target galaxies for this study are chosen to be two PRGs, which appear to be at two different evolutionary stages.  This is evidenced by the amount of structure and brightness of their polar ring. Both systems are located relatively nearby ($<\!100$\,Mpc) to facilitate reliable analysis of their most massive (brightest) star clusters, and imaging is available in the Hubble Space Telescope (HST) archive.

\begin{figure}[h]
\centering
\includegraphics[scale=0.335]{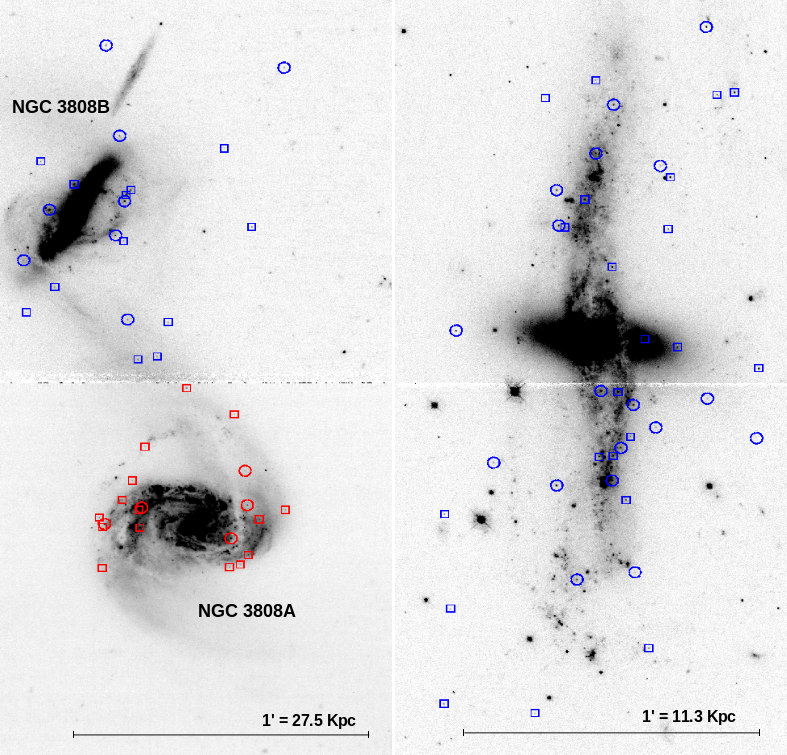}
\caption{F450W (B-band) HST/WFPC2 image of NGC\,3808 ({\bf left}) and NGC\,4650A ({\bf right}). Selected compact stellar system candidates are shown with symbols (see \S\,\ref{subsect: cluster selection}). Circles indicate the candidates detected in three filters, and squares are candidates detected in two filters ($F450W$ and $F555W$ or $F606W$).}
\label{fig:selection}
\end{figure}

NGC\,4650A is a massive polar disk galaxy (see Fig. \,\ref{fig:selection}, right). It is composed of a puffed up disk with no bulge as a central host galaxy \citep[HG;][]{Iodice02}, and a polar structure  that is rotationally supported and has young stellar population \citep[PRG catalog;][]{Whitmore90}. The polar ring has the characteristic of a disk \citep{arnaboldi97,Iodice02}.  \cite{Iodice02} perform a polygon photometry using optical and near-infrared data to estimate the age of the polar disk of $<\!0.5$\,Gyr and an age of 1\,--\,3\,Gyr for the host galaxy in comparison with simple stellar population (SSP) models. The metallicity of the polar disk is $Z\,=\,0.2\,Z_{\odot}$ ([M/H]\,$=\!-0.7$\,[dex]), and is derived from spectroscopic observations \citep{Spavone10}. The only previous study of the star cluster population of this galaxy is by \cite{Karataeva04_4650A}. Their analysis shows that these stellar systems follow the polar ring and their positions are correlated with HII regions. They derive very young ages for the stellar content of NGC\,4650A of up to 22\,Myr for an assumed $Z\,=\,0.4\,Z_{\odot}$, from the comparison of photometric colors to SSP model predictions from the Padova stellar evolution library.

NGC\,3808 is a double interacting system between two spiral galaxies (see Fig. \,\ref{fig:selection}, left). The main galaxy is NGC\,3808A (face-on) and its interacting companion is NGC\,3808B (nearly edge-on). The gravitational interaction of the two systems is evidenced by the tidal bridge via which gas is being transferred from the spiral arms of NGC\,3808A to NGC\,3808B. This gas transfer leads to the formation of a narrow ring around NGC\,3808B, which rotates on a nearly orthogonal plane \cite[spectroscopically confirmed by][]{Reshetnikov90}. The high star formation activity in this system is evidenced by the strong H$\alpha$ and infrared emission detected in NGC\,3808B \citep{Reshetnikov96}. They estimate an age for the polar ring of about 1\,Gyr.

The general properties of both targets are summarized in Table\,\ref{tab:properties}, along with the properties of the nearby more massive galaxy, NGC\,4650 and NGC\,3808A.

\input{table.tex}

\subsection{Data and image reduction}\label{Sect: Data and image reduction}
We retrieve the data used in this study from the HST science archive\footnote{http://hla.stsci.edu/}. The WFPC2 camera consists of four detectors, one high-resolution planetary camera (0.05$^{\prime\prime}$/pixel) and three wide field detectors (0.1$^{\prime\prime}$/pixel). NGC\,4650A and NGC\,3808 were observed as a part of the HST programs GO-8399 and GO-11092 (PI: Keith Noll), respectively.
The GO-8399 program observations are designed to study the star cluster population of NGC\,4650A. The GO-11092 is a Hubble Heritage program aiming at generating press release images of NGC\,3808.  Therefore, the exposure times of the latter are not ideally optimized to meet any specific criteria, nevertheless, as we show later, they are suitable for the goals of the current study.

The WFPC2 observations of NGC\,4650A are in three bands, $F450W, F606W, F814W$, with total integration time of 7700s, 4803s, and 7660s.  NGC\,3808 has observations in $F450W$ (9600s), $F555W$ (4400s) and $F814W$ (4400s).  The equivalent signal-to-noise (S/N) ratio estimated with the WFPC2 exposure time calculator (ETC) \footnote{http://www.stsci.edu/hst/wfpc2/software/wfpc2-etc.html} is approximately 80 and 20 within the PSF area, respectively, for an absolute magnitude of $M_V\!=\!-9.0$\,mag (see \S\,\ref{Sect: Completeness}). Therefore, the depth and resolution of these HST images are sufficient to detect and study the photometric and structural properties of the brightest stellar systems in these galaxies. One WFPC2 pixel on the wide field camera corresponds to 18.8\,pc at the distance of NGC\,4650A and 45.8\,pc for NGC\,3808. Hence, for objects with $S/N>30$ we can potentially measure sizes down to $10\%$ of the detector PSF, i.e., for WFPC2 $>1.8$pix (33.8pc for NGC\,3808 and 82.4pc for NGC\,4650\,A).

Retrieved archival images are processed by the HST pipeline, which corrects for flat field, bias, and dark. To remove high-energy cosmic rays from the images, we use the L.A. Cosmic IRAF procedure, {\sc lacosmic} \footnote{\href{http://www.astro.yale.edu/dokkum/lacosmic/}{http://www.astro.yale.edu/dokkum/lacosmic/}} \citep{VanDokkum01}.  This procedure uses a Laplacian edge detection algorithm to analyze the images and detect cosmic rays based on their sharp edge. For these highly under sampled HST images, we have carefully tested the {\sc lacosmic} parameters to avoid detection  by the algorithm of star or clusters tips as cosmic rays.

To register and combine the individual exposures into one final image with higher signal to noise, we use a custom developed procedure in IRAF.  This procedure is a wrapper for the {\sc geomap, geotran} and {\sc imcombine} IRAF procedures, and  it takes a few stars as reference in one frame to estimate the initial guess of the offsets between the reference and the rest of the images in the list. The procedure uses {\sc geomap} to compute the exact geometric transformation solutions and {\sc imcombine} is invoked for the final average combination of all registered exposures, which are brought to the same zero level as the reference evaluated from 30$\times$30 pixel statistic region away from the galaxies.

We calculate the distance modulus $m\!-\!M$ for our sample galaxies from the median value of the radial velocity measurements from all available and collected in HyperLEDA literature data. 
These values are listed in Table\,\ref{tab:properties}. Our $m\!-\!M$ differs from the {\tt modz} value in LEDA because we used the median of all $m\!-\!M$ values, while LEDA uses the mean, which is more prone to biases if there are outliers.

\subsection{Photometry}\label{Sect: Photometry}
Object detection in the images is performed with the IRAF task {\sc daophot/daofind}. To perform PSF photometry, we generate a PSF library with the {\sc TinyTim} software package \citep{Krist11}, which creates HST instrument tailored PSF models\footnote{TinyTim software accounts for variations in the PSF as a function of wavelength, position-dependent changes of the PSF shape, aberrations, filter pass-band, and telescope focus effects. For more details, http://www.stsci.edu/software/tinytim/tinytim.html}.  A grid of $10\!\times\!10$ PSFs for each WFPC2 detector are made and used as a PSF library to perform the photometry with the {\sc allstar} task. 

An important effect of the detector electronics, which needs to be accounted for, is the charge transfer efficiency (CTE). This can dim the magnitude by up to 0.2 mag for a low-count rate source and at low background levels. To correct for CTE, we use the latest correction coefficients provided together with the photometric zero points for the four WFPC2 detectors by \cite{Dolphin09}. Identically to the calibrations performed by the authors, we measure magnitudes within a radius of 0.5 arcsecond, i.e., an aperture radius of 5 WF pixels\footnote{No sources were detected on the high-resolution WFPC2/PC chip, therefore, all reduction and analysis are performed only for the WFPC2/WF detectors.}.  The same radius is used to estimate the aperture correction (AC) of the PSF magnitudes due to light loss from the finite size of the PSF model. We estimate the AC with artificial stars from the artificial star completeness test (see Sect. \ref{Sect: Completeness}), which allows us to cross-check with AC derived from field stars and obtain a more accurate value. PSF and aperture photometry of the artificial stars are performed with the same approach as for the science frames without artificial stars. To derive the AC, we use the difference between the PSF and 0.5$^{\prime\prime}$ aperture radius photometry.

Finally, we correct the photometry for foreground Galactic extinction. Extinction values for both galaxies are taken from the latest \cite{Schlafly11} recalibration of the \cite{Schlegel98} dust reddening maps.  Reddening values for the different filters are calculated assuming the \cite{Fitzpatrick99} reddening law with $R_V\!=\!3.1$.  The foreground extinction for NGC\,3808 are A$_{F450W}\!=\!0.089$\,mag, A$_{F555W}\!=\!0.072$\,mag, A$_{F814W}\!=\!0.04$\,mag and, for NGC\,4650A, are A$_{F450W}\!=\!0.388$\,mag, A$_{F606W}\!=\!0.274$\,mag, A$_{F814W}\!=\!0.176$\,mag.
 
\subsection{Completeness test}\label{Sect: Completeness}

Completeness tests are important for understanding the photometric depth of the observational data. We obtain completeness levels as a function of filter and position across the WFPC2 detector by iteratively adding artificial stars and recovering them with the same photometric techniques as for the science frames.

Artificial stars are added with the {\sc addstar} task in IRAF on the combined images for the different filters using the constructed PSF model used for photometry. The positions of the artificial stars are generated from a random seed value for the random number generator. We use the coordinates of one filter as a reference input to generate stars with realistic colors in the other filters based on the stellar models color range. To avoid image overcrowding we add 100 stars per image and repeat the process 100\,times, thus we add in total $10^4$ artificial stars per galaxy in each filter.
\begin{figure}
\centering
\includegraphics[trim=1.6cm 1.2cm 1.5cm 0cm, clip=true,totalheight=0.34\textheight]{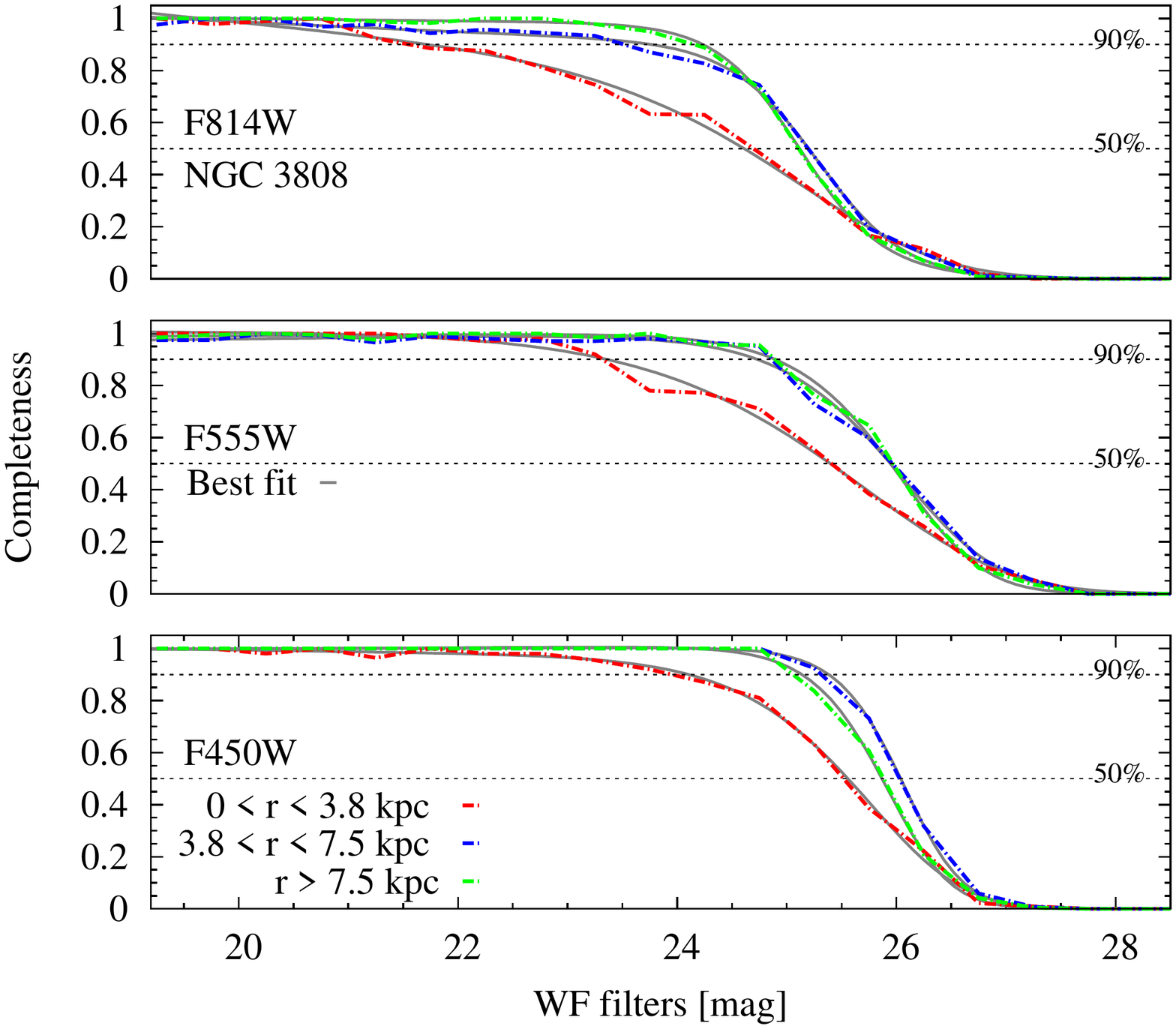}

\includegraphics[trim=1.6cm 5.7cm 0cm 5cm, clip=true, totalheight=0.167\textheight]{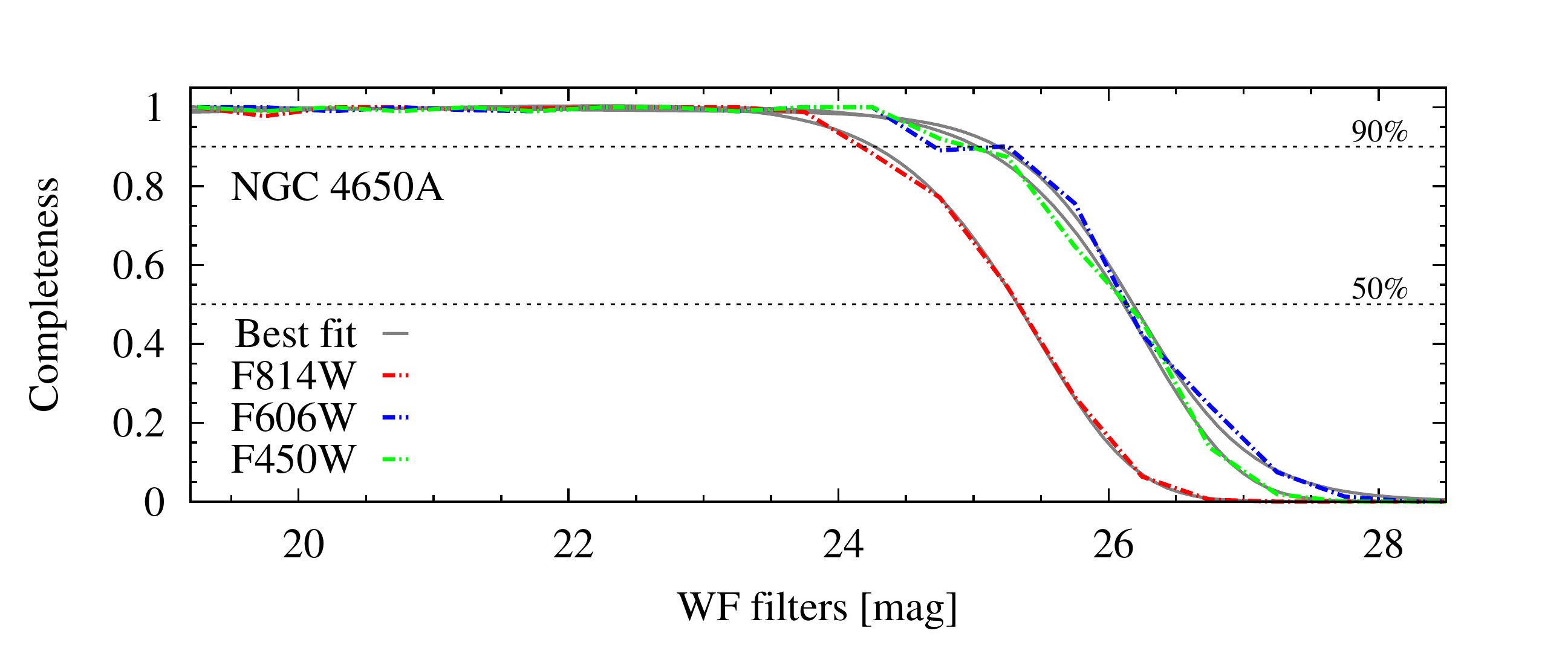}
\caption{ {\it Top three panels:} completeness functions from artificial star tests for the three different filters for NGC\,3808B. Each subpanel shows the completeness as a function of distance from the PRG's center in three annuli specified in the legend. {\it Bottom panel:} the overall completeness at all radii for NGC\,4650\,A is shown, where the individual radial incompleteness is nearly identical to the top three panels (see \S\,\ref{Sect: Completeness}). Dash-dotted horizontal lines indicate the $50\%$ and $90\%$ completeness levels. Solid curves show the power-law fits to the completeness results.}
\label{fig:comp}
\end{figure}
The same photometry procedure described in \S\,\ref{Sect: Photometry} is applied to these images (detection and photometry). The magnitudes of the artificial stars are then calibrated from instrumental to standard magnitudes using the same zero points and aperture corrections as for the science frames. The completeness is calculated as the ratio between the numbers of detected and input stars in a magnitude bin of 0.5\,mag. This is shown in Fig. \,\ref{fig:comp} for NGC\,3808, where we fit a function consisting of two power-law components to the completeness distributions. The completeness is tested as a function of the three WF filters (top to bottom panels), as well as a function of the distance from the galaxy center in three annuli, from 0 to $20\arcsec$, from $20\arcsec$ to $40\arcsec$, and beyond $40\arcsec$ (shown in each subpanel).  At the distance of NGC\,3808\,B, 20\arcsec is 3.8\,kpc and 40\arcsec is 7.5\,kpc. At 0.5 completeness, the bands $F450W$ and $F555W$ reach $\simeq26$\,mag, while $F814W$ is shallower by about 0.6\,mag.  This has implications for cluster selection in the three bands.  The completeness for the three different radial annuli shows that the incompleteness significantly increases by about 2\,mag toward the inner regions ($<3.8$\,kpc) of the galaxy because of the increasing surface brightness. This prevents us from detecting clusters fainter than about $M_V\!>\!-11$\,mag projected in the inner regions, whereas we have the deeper completeness ($M_V\!>\!-9$\,mag in the galaxy outer regions ($>3.8$\,kpc). The radial completeness functions in Fig. \,\ref{fig:comp} are shown only for NGC\,3808 and those for NGC\,4650\,A are nearly identical, where the incompleteness also drops by about 1-2\,mag toward the center with $F814W$ being the shallowest filter by $\simeq0.8$\,mag compared to the other two bands.

\subsection{Artificial star tests}\label{subsect: art_sharpness}

\begin{figure}
\centering
\includegraphics[trim=1.3cm 2cm 2cm 2cm, clip=true, totalheight=0.255\textheight]{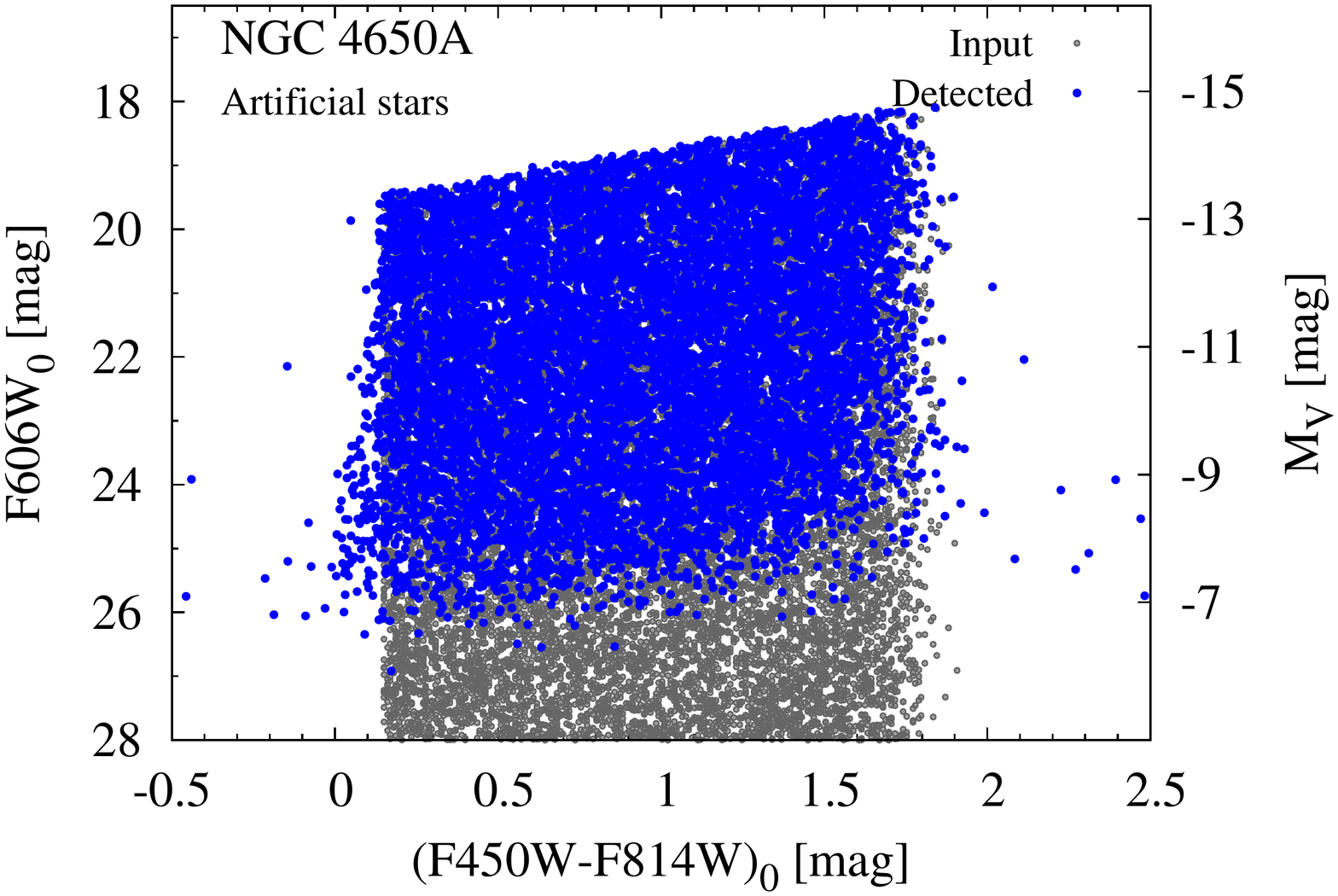}

\includegraphics[trim=1.5cm 2cm 1.5cm 2cm, clip=true, totalheight=0.25\textheight]{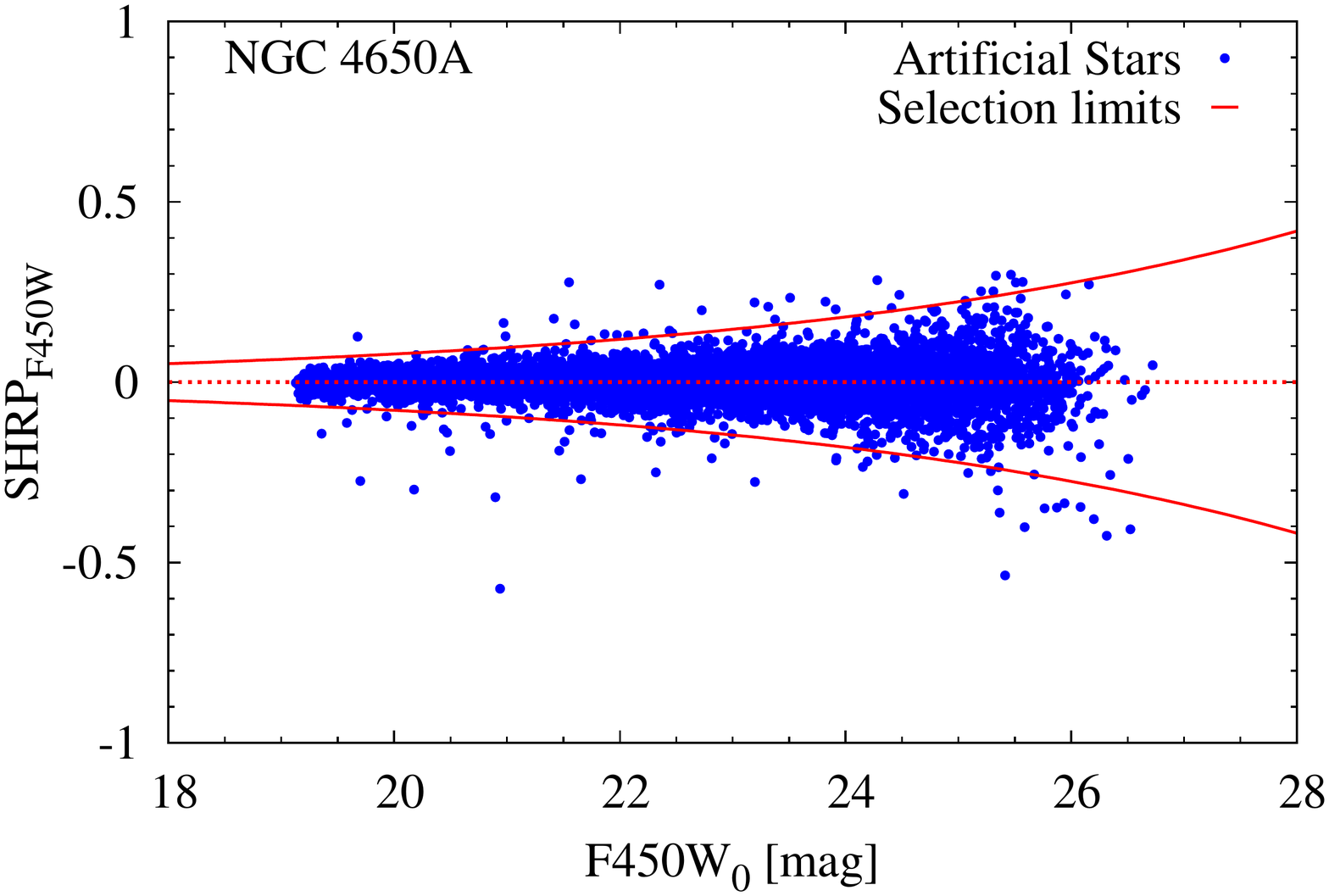}
\caption{Artificial star tests for the case of NGC\,4650A. The same analysis is performed for NGC\,3808. {\it Top panel:} color-magnitude diagram of input (gray dots) and recovered (dark blue solid dots) artificial stars. \textit{Bottom panel:} the artificial stars's sharpness value against their $F450W$-band magnitude. Red solid curve indicates the exponential cuts used for selecting unresolved objects (see \S\,\ref{subsect: cluster selection}).}
\label{fig:as_cmd}
\end{figure}

At the distance to the PRGs, star clusters are expected to be unresolved, i.e., with size comparable to or less than 10\% of the image PSF for $S/N\!>\!30$ sources. Only three objects had sufficiently high S/N to attempt to measure their size with {\sc ishape}, \citep{Larsen99}. Therefore, we use the artificial star photometry from \S\,\ref{Sect: Completeness} to test colors, magnitude, and sharpness values to optimize our selection criteria for unresolved objects. The output quality parameters from the PSF-fitting procedure {\sc allstar} are $\chi$ and {\sc sharp}. The $\chi$ is the classical goodness of fit parameter, while the sharpness value provides quantitative information about the difference between the object and stellar PSF profiles. For stars or unresolved objects, the sharpness value is close to zero.
Color-magnitude and magnitude against sharpness diagrams for the artificial stars are presented in Fig. \,\ref{fig:as_cmd}. The color-magnitude diagram (top panel) shows the input artificial stars (small, gray dots) and the detected stars (dark blue solid circles). From the artificial-stars CMD, it is evident that objects fainter than 26\,mag are no longer detected. The sharpness distribution (bottom panel) of the sources is very narrow for bright objects and gets broader toward the faintest magnitudes. To account for this behavior, for the sharpness-based selection of cluster candidates, we use an exponential function of the form: $f(x)= a\times{\rm exp}(\pm b \times x)$, where $a$ and $b$ are constants chosen to bracket the envelope of the artificial stars sharpness distribution (solid lines in Fig. \,\ref{fig:as_cmd}). This selection criteria for unresolved sources in the images utilizing the sharpness value helps to weed out a number of contaminating background galactic bulges, blemishes, and star-forming regions in the images (see also \S\,\ref{subsect: cluster selection}). Small contamination from foreground stars is expected in the color and magnitude range for the sky position of the WFPC2 field of view. We find that less than two stars per square arcmin are expected by using the Besan\c{c}on Galactic model\footnote{http://model.obs-besancon.fr/} \citep{Robin03}.

\subsection{Photometric metallicity and mass of the PRGs}\label{subsect: SSP}

In order to confine the color range for cluster candidate selection and subsequently their photometric age, we require prior knowledge of their likely metallicity. This can be closely approximated by the host PRG metallicity. Here we use single stellar population (SSP) models as the 2007 update to the \citep{Bruzual03}, hereafter CB07, to obtain a photometric constraint to the PRG metallicity, if not available from the literature measurement. With a canonical IMF \citep{Kroupa01} and Padova 1994 stellar evolutionary isochrones.  The CB07 models contain WFPC2 magnitude system. 
For the subsequent analysis, we also derive total photometric stellar mass of the galaxies using $M/L-$color relations \citep{Bell01,Bell03}.

The HST imaging data for both PRGs is only available in the optical and the known age-metallicity degeneracy prevents simultaneous and accurate calculation of both the age and metallicity of individual objects. For NGC\,4650A, we adopt the measured metallicity for the polar disk of Z\,=\,0.2\,Z$_\odot$ by \cite{Spavone10} and derive ages for that metallicity from the SSP models. For NGC\,3808, there are optical spectra available in SDSS for the nucleus of NGC\,3808\,B and for a bright HII region of the donor galaxy NGC\,3808\,A. These spectra contain strongly ionized gas emission lines that help us to derive the current gas abundance. We use the oxygen abundance methods followed in \cite{Spavone10}, such as the $R_{23}$ \citep{Pagel79} and the P-method \citep{Pilyugin01}, with $P=R_3/R_{23}$. The oxygen abundance $12+log(O/H)$ is estimated using Eq.\,2 in \cite{Spavone10}. For the nucleus of NGC\,3808\,B, the oxygen abundance is $12+log(O/H)=8.77$\,dex.  For the adopted $12+log(O/H)_{\odot}=8.83$\,dex \citep{Asplund04}, we obtain that the metallicity for the nucleus of NGC\,3808\,B is $Z\simeq0.87\,Z_\odot$.  For the HII region on the spiral arm of NGC\,3808\,A $12+log(O/H)=8.70$\,dex, which translates to a gas metallicity of $Z\simeq0.75\,Z_\odot$.

\begin{figure}
\centering
\includegraphics[trim=2cm 2.5cm 1.5cm 2cm, clip=true, totalheight=0.25\textheight]{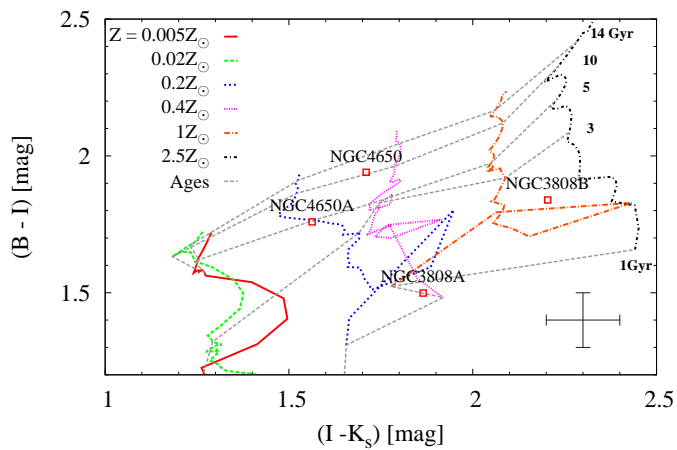}
\caption{Optical-NIR, color-color diagram, $B-I$ vs $I-K_S$ of the integrated colors of the PRGs and their neighbor galaxies shown with labeled red squares. Different gray dashed-lines show the age-labeled isochrones with increasing age from bottom to top for all metallicities in the \citeauthor{Bruzual03} SSP models as indicated in the figure legend. The typical photometric uncertainty of the galaxies' colors, given by SDSS and 2MASS, is about 0.2\,mag, and is shown in the bottom right.}
\label{fig:met_gal}
\end{figure}

To cross-check the metallicity estimate for the PRGs, we use the latest optical ($B, I$) and near-infrared (2MASS: $K_S$) data in VEGA magnitude available in the literature for these galaxies (cf. Table\,\ref{tab:properties}). The $I-$band magnitude for NGC\,3808 was transformed from the SDSS DR9 i-band photometry with the transformations in \cite{Chonis08}. The $B-I$ vs $I-Ks$ diagram is shown in Fig. \ref{fig:met_gal} for different ages (1, 3, 5, 10, 14\,Gyr) and metallicities ($Z/Z_{\odot}$ = 0.005, 0.02, 0.2, 0.4, 1.0, 2.5).  Red squares indicate the location of  the PRGs and their closest neighbor galaxies. NGC\,4650A metallicity ($Z\!=\!0.2\,Z_{\odot}$) agrees with the mean value derived spectroscopically by \cite{Spavone10}, which is only representative for the metallicity of the polar disk and where the majority of the young stellar systems are located. Its bright neighbor NGC\,4650 has a metallicity of about 0.4\,Z$_{\odot}$. Keeping the photometric uncertainties in mind, NGC\,3808\,B appears to be more metal rich ($Z\!=\!1\!-\!2.5\,Z_{\odot}$) compared to the metallicity derived from SDSS spectra ($Z\!=0.87\,Z_{\odot}$). For the PR NGC\,3808\,B, we adopt a mean metallicity of $Z\!=1\,Z_{\odot}$.  Its companion, NGC\,3808A, has a metallicity in the range of $Z\!=\!0.4\!-\!1\,Z_{\odot}$, which agrees with the $Z\,=0.75\,Z_{\odot}$ derived from SDSS spectra. We assume a mean metallicity for NGC\,3808\,A of $Z\!=0.4\,Z_{\odot}$.  However, the very low age for this galaxy may be explained by uncertainties in the apparent $B-$band magnitude. Integrated magnitudes of the PRGs and the galaxies are prone to internal reddening and inclination uncertainties, which we do not attempt to account for. These are less important in the NIR. Therefore, the integrated galaxy color analysis enables us to adopt a proxy metallicity for deriving cluster ages in NGC\,3808B.

To calculate the total galaxy stellar mass, ${\cal M}_{\star,\rm gal}$, we use the apparent galaxy magnitude from LEDA as well as $u,z,i$ SDSS magnitudes for the northern PRG NGC\,3808A,B. For NGC\,4650A, we adopt the total apparent magnitudes measured by \cite{Gallagher02}.
To estimate the galaxies' masses, we use the  $M/L-$color relations, which are empirically constrained from a suit of galaxy evolution models for a range of galaxy star formation histories \citep{Bell01,Bell03}. The derived ${\cal M}_{\star,\rm gal}$ are tabulated in Table\,\ref{tab:properties}. In general, in both cases, the PRG and the nearest donor galaxy have nearly comparable stellar masses. In the case of the NGC\,3808 system, the galaxies have a lower mass, ${\cal M}_{\star,\rm gal}\geq10^9M_\odot$, compared to NGC\,4650A and NGC\,4650, which are an order of magnitude more massive, ${\cal M}_{\star,\rm gal}\geq10^{10}M_\odot$. We discuss the mass ratios in \S\,\ref{Sect: Discussion} in the context of PRG formation.

\subsection{Cluster selection criteria}\label{subsect: cluster selection}
To detect and study the compact stellar systems, we employ information about the roundness of the detected sources. This allows us to differentiate between unresolved star clusters at the distance to the studied PRGs from other objects in the field of view, e.g., extended star complexes and background galaxies. The sharpness value is a proxy for object's roundness as it is defined as the difference between the object and PSF FWHMs. The chosen functional selection limits as a function of object magnitude (see Fig. \,\ref{fig:as_cmd} bottom panel) allows us to include sources with a slightly more extended light profile than the stellar PSF, which might be the case for the largest globular clusters and ultra compact dwarf galaxies (UCDs); \cite[$>\!20$\,pc, e.g.;][]{Misgeld11}.
To additionally help with avoiding background galaxies, we impose color selection limits at $F606W-F814W\!<\!1.0$\,mag and $F450W-F606W\!<\!1.5$\,mag.

The applied selection criteria can be summarized as follows: First we select objects that are detected in the three filters with: ($i$) a distance from the SSP model for an assumed metallicity in the two-color parameter space of less than $3\sigma$ away from the SSP model in the color-color space. The adopted distance from the SSP is comparable to the photometric uncertainty, $\sigma_{\rm dist}$=\,0.227\,mag (NGC\,3808) and $\sigma_{\rm dist}$=\,0.315\,mag (NGC\,4650A); ($ii$) color error $\leq\!0.2$\,mag and ($iii$) round and unresolved sources in all filters, using the exponential limits obtained from the artificial star sharpness distribution (see \S\,\ref{subsect: art_sharpness}). Because of the age-metallicity degeneracy, the SSP tracks are parallel (cf. \S\,\ref{fig:cc}) and nearly overlapping, and thus do not have a major effect on the cluster selection as is the case for their mass determination (see discussion in \S\,\ref{Sect:CSSs ages and masses}).

To maximize cluster detection due to variable completeness in different filters, for a second selection we use objects that are detected in two filters. We use the two deepest filters according to the completeness test, which are $F450W$ and $F555W$ or $F606W$. In addition to the two-color selection, we use criteria ($ii$) and ($iii$) for selecting CSSs. The final selected cluster candidates are shown in Fig. \,\ref{fig:selection}. 

\section{Analysis}\label{Sect:Analysis}
\subsection{Color-magnitude and color-color diagrams}\label{Sect: CMD and CCDs}

\begin{figure}
\centering
\includegraphics[trim=2cm 2cm 1.5cm 2cm, clip=true, totalheight=0.25\textheight]{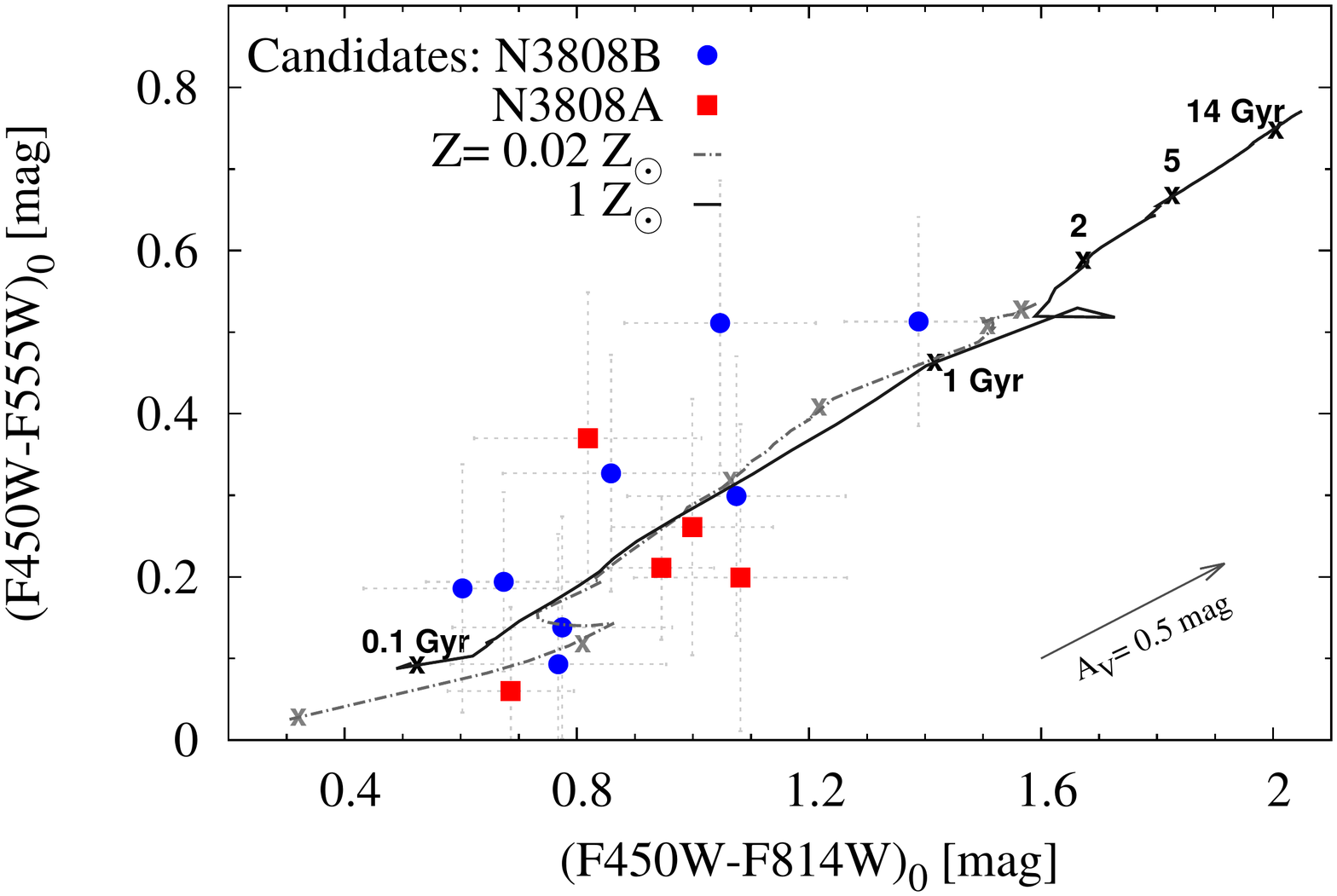}
\includegraphics[trim=2cm 2cm 1.5cm 2cm, clip=true, totalheight=0.25\textheight]{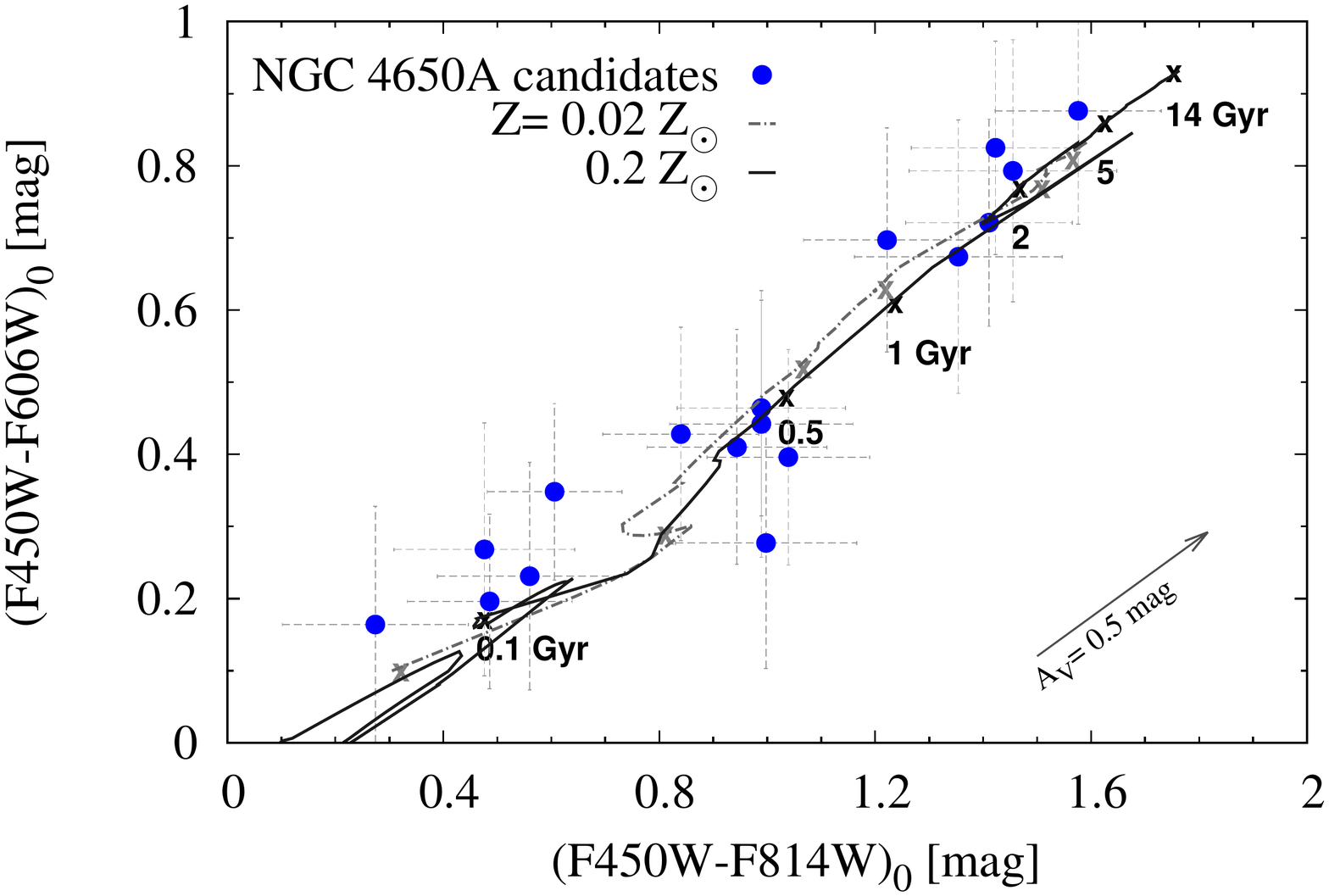}
\caption{Color-color diagrams in the $F450W - F555W/F606W$ vs. $F450W - F814W$ filters of the cluster candidates in both galaxies. {\it Top panel}: the NGC\,3808B stellar systems (the PRG) are shown with circles and the candidates in NGC\,3808A are indicated with squares (the gas donor galaxy). The assumed metallicity (see \S\,\ref{subsect: SSP}) for NGC\,3808B,  $Z=1\,Z_\odot$, is shown with a solid line. {\it Bottom panel}: NGC\,4650A cluster candidates are shown with blue solid circles, and the isometallicity track ($Z=0.2\,Z_\odot$) is plotted with a solid line. For comparison, we show with dashed line a metal-poor SSP of $Z=0.02\,Z_\odot$, typical for an old cluster. We indicate a few ages along the SSP tracks with labeled crosses.}
\label{fig:cc}
\end{figure}

In this section, we analyze the color-color and color-magnitude distributions of the selected CSS candidates. Color-color diagrams for the CSS candidates in both galaxies are shown in Fig. \,\ref{fig:cc}. 
The CB07 isometallicity tracks are shown with solid lines for ages between 100\,Myr to 14\,Gyr.  For the case of NGC\,3808, we show the metallicity used for the PRG, $1\,Z_\odot$ derived as described in \S\,\ref{subsect: SSP} and for comparison a metal poor track of $0.02\,Z_odot$.  A common feature observed for both galaxies is that most of the candidates have blue colors, indicating young ages (most with age $<$\,1Gyr).  The color distribution of NGC\,3808 objects is concentrated around colors $0.4\!<\!F450W-F814W<1.2$\,mag, while NGC\,4650A objects show a larger spread, extending to a redder $F450W-F814W$ color. These color distributions suggest a large spread in age (if all clusters have the same metallicity) or vice versa, a large spread in metallicity (if all clusters have comparably similar age and formed during the same event). The effect of unknown internal reddening further add to the uncertainty, and is shown with an arrow in Fig. \,\ref{fig:cc}.

\begin{figure}
\centering
\includegraphics[trim=2cm 2cm 1.5cm 2cm, clip=true, totalheight=0.26\textheight]{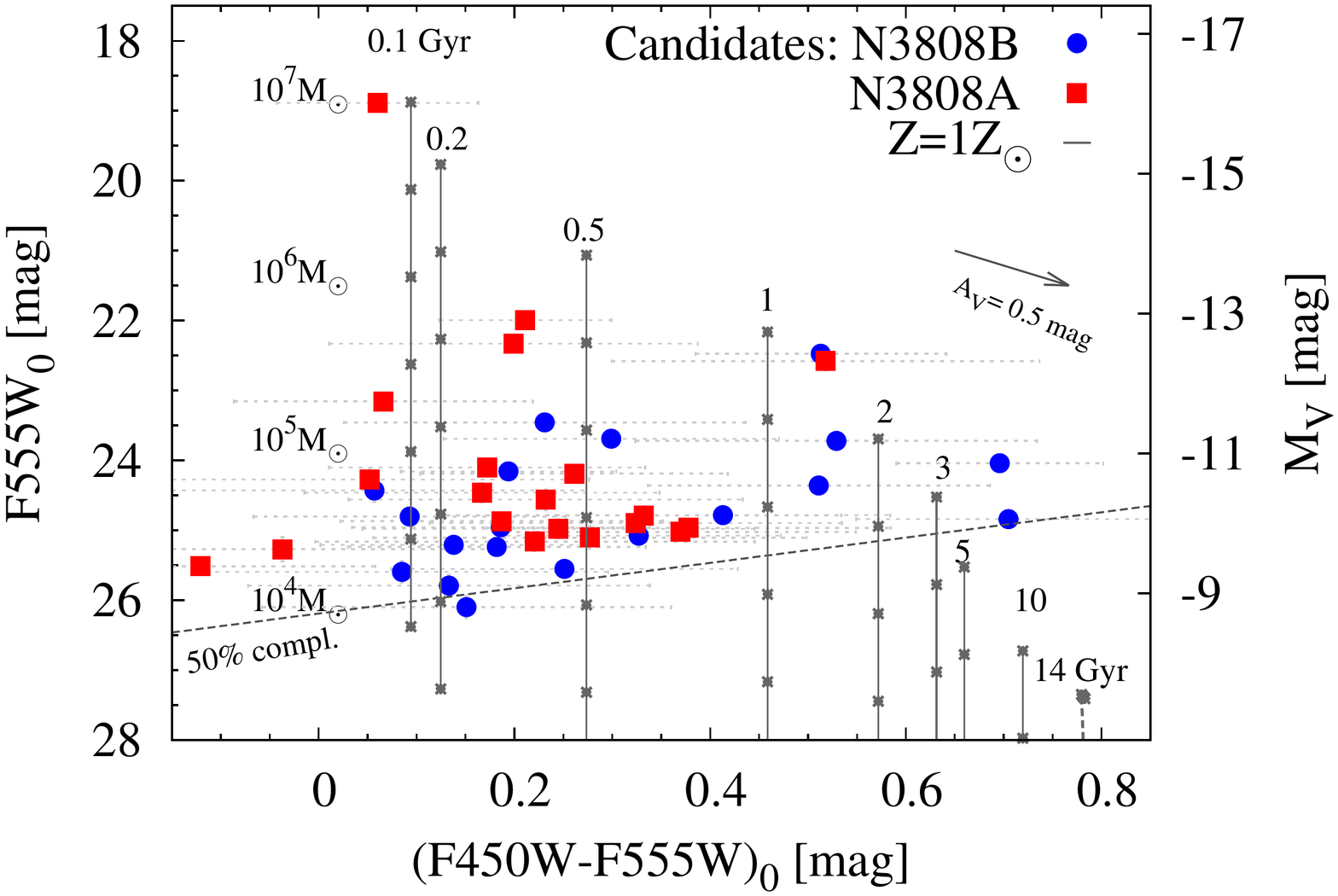}\\
\includegraphics[trim=2cm 2cm 1.5cm 2cm, clip=true, totalheight=0.26\textheight]{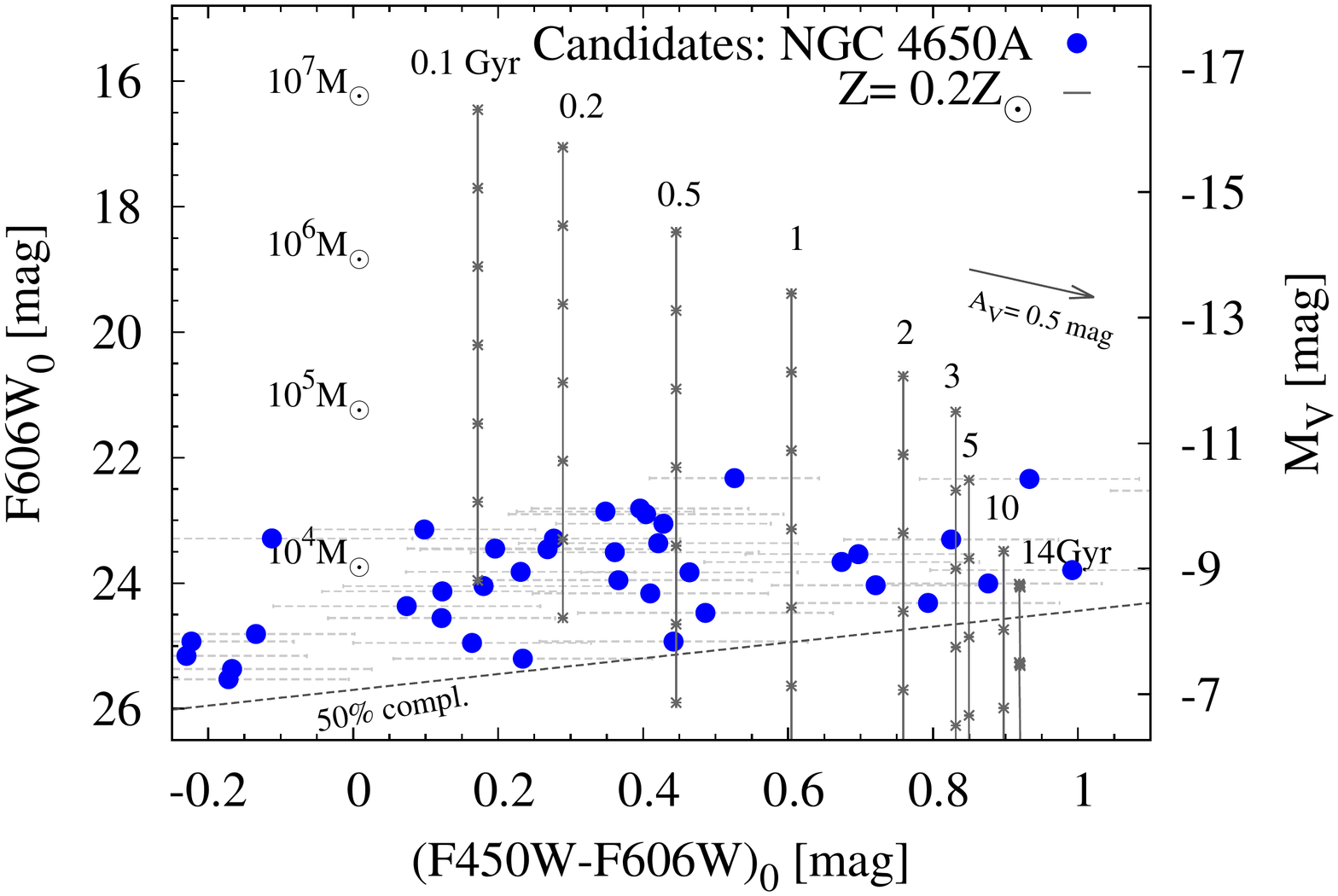}
\caption{Color-magnitude distribution of all sources in NGC\,3808 (top) and NGC\,4650A (bottom). Vertical isochrone lines for ages from 0.1 to 14\,Gyr are shown with solid lines for a fixed metallicity, $Z=1Z_\odot$ for NGC\,3808 and $Z=0.2Z_\odot$ for NGC\,4650A. Along the vertical lines the masses are labeled at a given age. Completeness at 50\% is shown with dashed line. The arrow indicates a reddening of 0.5\,mag.}
\label{fig:cm}
\end{figure}

The color-magnitude distribution of the CSSs in the two PRGs is presented in Fig. \,\ref{fig:cm}. For reference, isochrones are shown with vertical solid lines for ages ranging from 100\,Myr to 14\,Gyr from left to right. Along the isochrones, asterisk symbols indicate the masses from $10^4$ to $10^7M_\odot$. Although we have constrained the possible range in metallicity, the intrinsic galactic reddening remains uncertain.  The magnitude of this effect can be seen by the arrow in the color-color plots.  Further analysis in terms of age and mass for objects is given in the next section. 
We find that NGC\,4650A has stellar systems with $F606W$ (V) between 25.8 and 22 mag, i.e, $-11<\,M_V\,<\,-7$\,mag at the distance of the galaxy. These values are in the typical luminosity range of bright star clusters and in the low-luminosity range of UCDs. We discuss in detail the most massive and brightest UCD candidate, NGC\,3808\,B-8, in \S\,\ref{Sect:Discussion:ClusterPops}. The majority of these systems are located along the polar disk. This is in agreement with the magnitudes for the objects found by \cite{Karataeva04_4650A} in NGC\,4650A. 

On the other hand, NGC\,3808 contains bright CSSs, most of which have absolute magnitude around $-13<M_V<-9$\,mag. The brightest objects in Fig. \,\ref{fig:cc}, with $M_V\simeq-13$\,mag, are located in the spiral arms of NGC\,3808A and one object is located on the polar ring of NGC\,3808B.  
CSSs with luminosities brighter than $M_V=-10$\,mag are good candidates for young globular clusters and UCDs. As this is the first study investigating the star cluster systems of NGC\,3808, we cannot perform a comparison with the literature.  We find that the location of the CSSs is mostly in the tidal bridge connecting the two galaxies, on the disk of NGC\,3808A, and along the polar ring of NGC\,3808B. Candidates along the disk may suffer from internal galactic reddening (see reddening vector in Fig. \ref{fig:cc}).

\subsection{Photometric star cluster age and mass estimates}\label{Sect:CSSs ages and masses}

Photometric ages of the clusters are estimated from interpolation between object and SSP model colors using a fixed metallicity; for the assumed metallicities we refer to Sect. \S\,\ref{subsect: SSP}. The uncertainty in the derived ages are estimated based on the propagation of the photometric uncertainty.  We perform a Monte-Carlo simulation by drawing 100 random colors from a normal probability distribution function with a mean corresponding to the observed color and a standard deviation from the observational color error. The most robust age estimate at fixed metallicity is for objects that are detected in all three filters, i.e., two-color information is available. Otherwise, age is only obtained from one color $F450W-F555W$ or $F450W-F606W$ color, $(B-V)$. In Fig. \ref{fig:cm}, we show the color-magnitude diagram of cluster candidates detected in the two deepest filters (c.f.\S\,\ref{Sect: Completeness}), which yields a larger number of sources compared to Fig. \ref{fig:cc}. We have corrected magnitudes for Galactic foreground extinction, but no corrections for intrinsic absorption from the galaxies themselves. Thus, the derived here ages should be considered an upper estimate. 
Because the main aim of this paper is to study the star cluster population associated with the younger, polar-ring structures, it is probably fair enough that we assumed a SSP model with the same metallicity as the PR, i.e., metal-rich (see discussion in \S\,\ref{Sect:degeneracy}).

\begin{figure}
\centering
\includegraphics[trim=2cm 2cm 1.5cm 2cm, clip=true, totalheight=0.26\textheight]{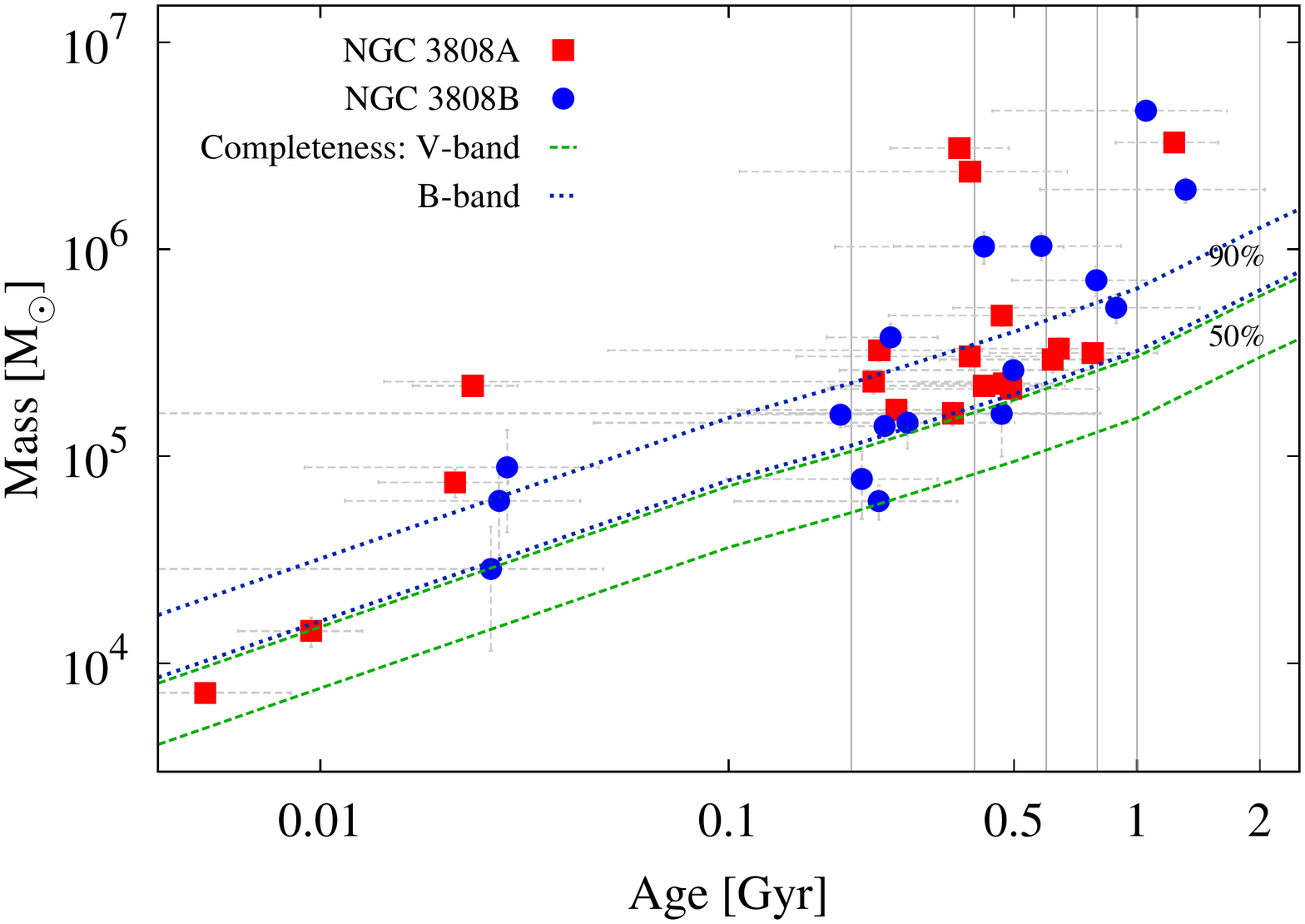}
\includegraphics[trim=2cm 2cm 1.5cm 2cm, clip=true, totalheight=0.26\textheight]{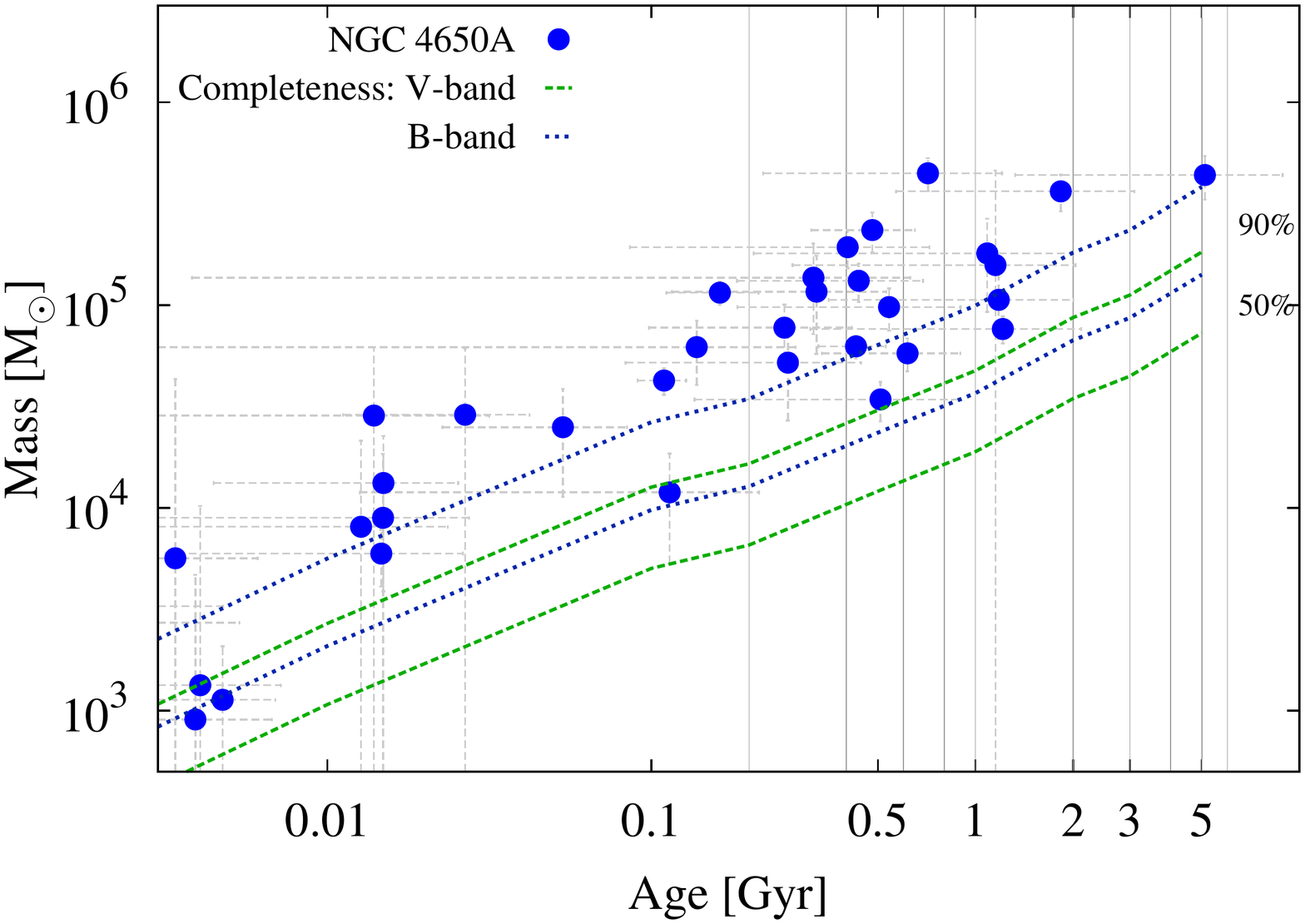}
\caption[Mass vs age for the CSSs of the two PRGs.]{Mass versus age for the cluster candidates in the two PRGs, NGC\,3808 (top), and NGC\,4650A (bottom). Red squares are for NGC\,3808A objects and blue circles for NGC\,3808B. The completeness curves at 50\% and 90\% for the $V$- and $B$-bands are shown with dashed and dotted lines, respectively. Vertical lines show the age bins used to estimate the SFR; see \S\,\ref{sec:sfrprg}.}
\label{fig:age_mass}
\end{figure}

We calculate the mass of the compact stellar systems using SSP $M/L_V$ ratio from CB07 models using object's age for an assumed metallicity. We use the luminosities of the clusters calculated using the distance modulus to the galaxies as given in Table\, \ref{tab:properties}.  The distribution of the derived CSSs' age and mass for both galaxies is shown in Fig. \,\ref{fig:age_mass}. 
Completeness limits at 50\% and 90\% are indicated and are calculated by converting the completeness function in $V$ band with the SSP $M/L_V$ ratio at different ages.  This is used to determine the mass completeness function at an assumed metallicity, $Z=\,1Z_\odot$ for NGC\,3808 and $Z=\,0.2Z_\odot$ for NGC\,4650A. The incompleteness increases for the oldest and intermediate mass clusters, as can be seen seen in Fig. \,\ref{fig:age_mass}.

Both galaxies show a young star cluster population, with most of them consistent with ages $<1$\,Gyr. For NGC\,3808, the ages of the tidal bridge objects are $<400$\,Myr and up to 1.5\,Gyr for stellar systems around the polar ring in NGC\,3808B.  \cite{Reshetnikov96}, using integrated colors of the polar ring of NGC\,3808B, estimate its age to be 1\,Gyr, which is consistent with our results.

For NGC\,4650A, the age distribution shows that more than 80\% of its detected sources are significantly younger than 1\,Gyr with most of the objects located along the polar disk. A few sources have colors consistent with ages greater than 1 Gyr. Their spatial location is close to the central component or in the outer part of the polar ring. If they belong to the older host galaxy population their ages would be comparable to the range derived previously for this component, which is $\simeq3-5$\,Gyr \citep{Iodice02,Gallagher02}. The latter age is a luminosity-weighted age estimate, hence a more luminous younger population would bias toward overall younger age.  Given the unknown extinction, the older clusters may be reddened genuine older globular clusters. However, Fig. \,\ref{fig:age_mass} (cf. 90\% completeness limit), shows that we are significantly incomplete in the detection of the oldest objects($>10$\,Gyr), such as globular clusters with a typical luminosity of $M_V=-7.5$\,mag.

We find that the NGC\,3808 CSSs have masses from $10^4M_\odot$ to $5\times10^6M_\odot$.  NGC\,4650A contains objects within a mass range of $10^3M_\odot$ to $5\times10^5M_\odot$.  Stellar systems with masses below $10^4M_\odot$ are very young, which can also be contaminant supergiant stars and very young super star clusters. \cite{Karataeva04_4650A} assumed that all such bright sources detected in NGC\,4650\,A are blue/red supergiant stars. However, their estimate may be affected by a large number of background sources because of their very broad range of sharpness($-1< {\rm SHARP} <1$). As we showed with the artificial star test in \S\,\ref{subsect: cluster selection} (Fig.\,\ref{fig:as_cmd}), unresolved, point sources have a sharpness value of up to $\pm0.3$. Therefore, their selection may contain a significant number of contaminating stellar associations and blemishes in the star-forming regions, as well as background galaxies, which can give rise to their high number of blue objects. Our narrower sharpness selection guarantees that the majority of these are weeded out.\\
We can safely assume that most of the brightest ($M_V\!<\,-\!10$\,mag) are genuine star cluster candidates because this luminosity implies high-mass blue and red supergiant stars.  These stars are very short-lived, normally a few Myrs, and according to the comparison with the models, most of our bright sources are likely older than 50\,Myr, thus older than the life of a supergiant star.
However, some of the CSSs are projected onto HII regions. Given the relatively low surface brightness of these regions, compared to that of the cluster, we do not expect that strong emission lines (e.g., from $H_\alpha$) contained in $F606W$ would strongly contribute to the flux to add a systematic uncertainty to our age-metallicity estimates.

\subsection{Caveats due to age-metallicity degeneracy}\label{Sect:degeneracy}

As mentioned in \S\,\ref{subsect: SSP}, the choice of metallicity does not affect our cluster selection because the age-metallicity degeneracy makes the SSP model trace parallel in the color-color diagram (see \S\,\ref{subsect: cluster selection} and Fig.\,\ref{fig:cc}).
Using the same metallicity as the PRGs for deriving the cluster properties may be a fair assumption to study the clusters that formed and are associated with the PRs, however, this choice of metallicity may not be appropriate for the oldest and metal-poor cluster. As a general test to assess the age uncertainties due to the choice of metallicity, we estimate the age of the cluster candidates using the most metal-poor (0.005\,Z$_\odot$) and the most metal-rich (2.5\,Z$_\odot$) SSPs in the CB07 models. The difference between the clusters' age obtained from assuming the metallicity of the PRG and that for a metal poor/rich SSPs is between $\Delta t=1.5$\,Gyr for the most metal-poor to $\Delta t=0.25$\,Gyr for the most metal-rich tracks.

We check whether some of the detected clusters could be genuine old GCs by calculating their ages with a low metallicity track, $Z=0.02Z_\odot$ (i.e., LMC metallicity) and examine the location of the oldest candidates. We find that there are five candidates in NGC\,3808\,B that have ages older than 2\,Gyr. The only candidate with three-filter (two-color) photometry is the most massive cluster candidate, which is also located close to the PRG and its age is 2\,Gyr for $Z=0.02\,Z_{\odot}$ compared to an age of 1\,Gyr for $Z=1\,Z_{\odot}$. Unfortunately, the ages of the rest of the clusters are only calculated with two-band photometry ($B$ and $V$, but not detected in $I$ due to incompleteness). Thus, their age is calculated from only one color, which makes it highly uncertain due to the age-metallicity degeneracy. Three of them are spatially projected close to the polar ring, including the most massive cluster in NGC\,3808\,B, which may suggest that they are more likely to be metal-rich, bright young clusters. These could be even brighter if we also account for the internal reddening in those regions, which is unknown. There are only two candidates located further from the PRG, which have an age of about 7-8\,Gyr. However, these ages are unreliable because are calculated using only one color.

There are six clusters in NGC\,4650\,A with ages greater than 3\,Gyr and all with two color information. Four of them have ages younger than 6\,Gyr and two are older than 8\,Gyr. However, the latter are also located close to the polar ring instead of further out in the galaxy halo, as would be expected for old GCs.

Therefore, we are not really confident that we are indeed detecting some of the oldest, metal-poor halo GCs. In addition the HST/WFPC2 fields have a very limited coverage of the galaxies' outer regions. These clusters, however, could be considered intermediate-age cluster candidates, whose more precise age and metallicity estimate will await a deeper optical-NIR follow-up observations. Nevertheless, the general discussion and conclusion does not change significantly due to the choice of SSP metallicity, beyond the fact that 4-8\,Gyr old GCs could have formed during an interaction event 7-8\,Gyrs ago out of relatively unpolluted metal-poor gas. Because our focus is on the analysis of the star cluster population of the polar rings, we focus our discussion on the cluster properties' distributions for similarly high metallicity as the PRGs, as obtained in \S\,\ref{Sect:CSSs ages and masses}.

\subsection{Estimating the star formation rates of the PRGs}\label{Sect:PRG SFHs}
\label{sec:sfrprg}

There are number of ways of assessing the SFH of galaxies. To gain insight into the SFH and assembly of our sample PRGs, we calculate the SFR via the relation between the mass of the most massive cluster and the SFR at a given epoch \citep{Larsen02,Weidner04}. As mentioned in \S\,\ref{Sect:Intro}, this empirically calibrated relation reflects the fact that intensive SFR leads to the formation of a higher number as well as to more massive clusters. The method has been successfully tested in deriving the LMC SFH \cite[cf.\,\S\,\ref{Sect:Intro};][]{Maschberger11} and applied to other massive galaxies \citep{Bastian08,Georgiev12}. 

\begin{figure}
\centering
\includegraphics[width=.5\textwidth, bb = 2.2cm 4cm 25.9cm 18.5cm]{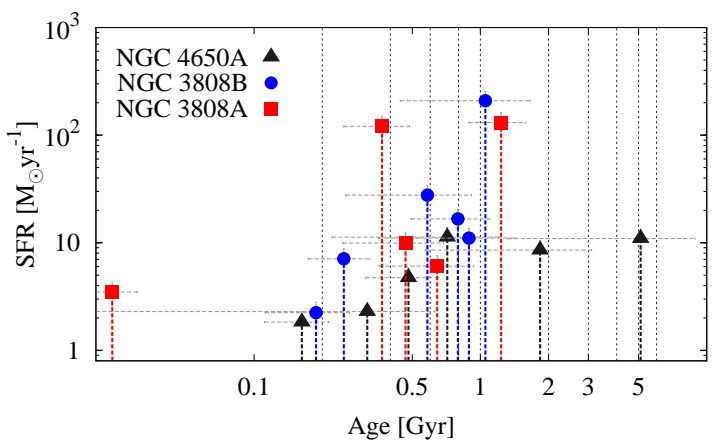}\\
\caption{Star formation rate vs. the age of the most massive cluster within the respective age bin indicated with light vertical dashed lines. The interacting system NGC\,3808 is shown with red squares (donor galaxy) and blue circles (PRG), respectively.  Black triangles indicate the PRG NGC\,4650\,A. 
}\label{fig:sfr}
\end{figure}

We reconstruct the SFH of NGC\,3808 and NGC\,4650A following similar approaches as in \cite{Bastian08,Georgiev12}.   
From the photometrically derived masses of the CSSs, we select the most massive objects in age bins of 200\,Myr for ages younger than 0.9\,Gyr and 1\,Gyr bin for ages older than 0.9\,Gyr for both PRGs. These age bins are shown with vertical lines in Fig. \,\ref{fig:age_mass} and \ref{fig:sfr} and are chosen to be comparable to the uncertainties of the derived ages.  We use Eq.\,7 in \cite{Weidner04} to estimate the SFR from the mass of the most massive cluster in those age bins. The mass loss due to stellar evolution was calculated from the $M/L_V$ values at 10\,Myr with \cite{Bruzual03} SSP models.  We did not attempt to correct for dynamical mass loss because we lack exact knowledge on the orbital parameters of the clusters. That defines the efficiency of tidal shocks on cluster mass loss due to encounters with giant molecular clouds, shocks from crossing the galactic disc and/or close passages to the nucleus/bulge depending on orbit eccentricity.  Relaxation  driven mass loss, on the other hand, is not expected~to~be~too~sig\-nificant for massive clusters. This effect has a dependence of mass described by $d(M/M_\odot)/dt = - (M/M_\odot) ^{1-\gamma/t_0}$ \citep{Baumgardt03,lamers10}. The index $\gamma$ varies from 0.65-0.8 according to N-body simulations, depending on the initial density distribution, and $t_0$ is the dissolution parameter that depends on the environment.

In Fig. \,\ref{fig:sfr} we show the derived SFRs as a function of the age of the most massive cluster in a given age bin. 
NGC\,3808A has two well-pronounced peaks of star formation at around 1\,Gyr ($\simeq130M_\odot$/yr) and 400\,Myr ($\simeq120M_\odot$/yr).  The strongest peak of SFR at around 1-2\,Gyr also corresponds with the strongest peak of SFR in NGC\,3808B (SFR\,$\simeq208M_\odot$/yr). The temporal coincidence of the two peaks is in support of the tidal interaction triggered star formation. In addition, the most massive star cluster in NGC\,3808\,B, with an age of 1-2 Gyr, is located in the polar ring.  A similar estimate of 1\,Gyr for the age of the polar ring is obtained by \cite{Reshetnikov96} from the analysis of optical colors for solar metallicity.\\
To obtain an independent estimate of the current SFR of these galaxy pairs, we searched for available data in a broader wavelength range, for instance $H_\alpha$ and FUV.  The $H_\alpha$ flux were measured by \cite{Gavazzi06}. For the PR NGC\,3808\,B, the  $L_{H_\alpha}=5.9\times10^{41}\,erg/s$, using the SFR$-{H_\alpha}$ relation \citep{Hunter04} and $A_{H_\alpha}=0.811\,A_V$, we calculate a present-day SFR$_{H_\alpha}\simeq$\,3.7\,M$_\odot$/yr. For the donor galaxy NGC\,3808\,A, the $L_{H_\alpha}=8.5\times10^{41}\,erg/s$, thus it has a current SFR$_{H_\alpha}\simeq$\,5.3\,M$_\odot$/yr. For available NED data in FUV, we use \cite{Kennicutt98} to derive the current SFR for NGC\,3808\,B of SFR$_{FUV}=0.29M_\odot$/yr for L$_{FUV}=2.07\times10^{27}\,erg/s$. For NGC\,3808\,A we find SFR$_{FUV}=1.27\,M_\odot$/yr.  This shows a similar trend as obtained from $H_\alpha$, i.e., the donor galaxy has a higher SFR than that of the PR galaxy.  These SFR values are higher than the typical present-day SFR values of starburst dIrr galaxies \cite[e.g.;][]{Hunter04,Georgiev06}, confirming the tidally triggered mode of star formation.

The SFR history of NGC\,4650A is measured up to a lookback time of 5\,Gyr because of the strong incompleteness at older ages (cf. Fig. \,\ref{fig:age_mass}). The peak value of the SFR is $<\!10M_\odot$/yr. This incompleteness comes from the photometric depth of our data and the limited spatial covering of the HST/WFPC2, which hampers the detection of lower mass/older clusters in the PRs, e.g., along the host galaxy's halo component.  Therefore, with the current data we cannot directly probe the SFR at the very first epoch of star formation, which triggered the formation of NGC\,4650A PR, similar to our approach for NGC\,3808B. Deeper and multiwavelength spectroscopic and imaging observations are required to probe the earliest epochs of star formation of NGC\,4650A.  The current SFR for the polar ring of NGC\,4650\,A, obtained from $H_\alpha$ luminosity, is SFR=$0.06M_\odot$/yr \citep{Spavone10}. The available FUV measurement in NED for NGC\,4650\,A is L$_{FUV}=4.93\times10^{25}\,erg/s$, yielding SFR$_{FUV}=0.007M_\odot$/yr. The nearby spiral galaxy (NGC\,4650) has L$_{FUV}=1.56\times10^{26}\,erg/s$, yielding SFR$_{FUV}=0.022\,M_\odot$/yr. There is a difference in the FUV-based SFR estimates, where the current SFR of the PRG NGC\,4650\,A is 30\% that of the NGC\,4650. If we normalize to the total, galaxy baryon mass (stellar plus HI) translates to a specific SFR (sSFR) of $1.56\times10^{-13}yr^{-1}$ for NGC\,4650\,A and $3.93\times10^{-14}yr^{-1}$ for its companion.  Therefore, we find that the sSFR is larger for the polar disk galaxy than for NGC\,4650.

In general, we observe that the earliest epoch of star formation in both galaxies attained the strongest SFRs. Subsequently, more recent star formation episodes proceeded with a lower SFR, reflected in the formation of clusters with lower mass. The low present-day SFR is confirmed, for the case of NGC\,3808B, from the SFR calculated based on $H_\alpha$ flux. We discuss in \S\,\ref{Sect: Discussion} the possible implications for the formation channels of the PRGs.

\section{Discussion}\label{Sect: Discussion}

In the following, we discuss the implications for the SFH and formation channels of the PRGs, based on their star cluster system properties.

\subsection{Star cluster populations and evolutionary stage of the PRGs}\label{Sect:Discussion:ClusterPops}

\begin{figure}
\centering
\includegraphics[width=0.47\textwidth]{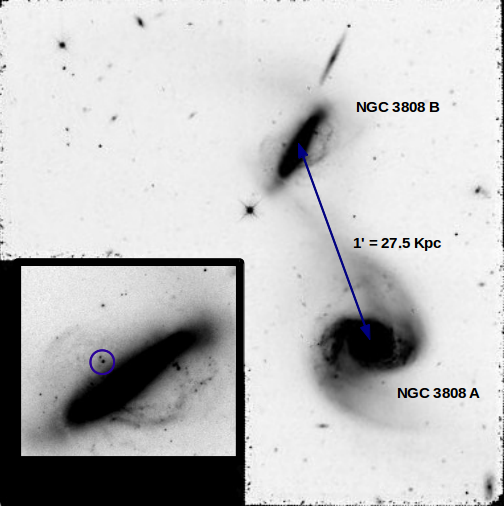}\\
\includegraphics[width=0.47\textwidth]{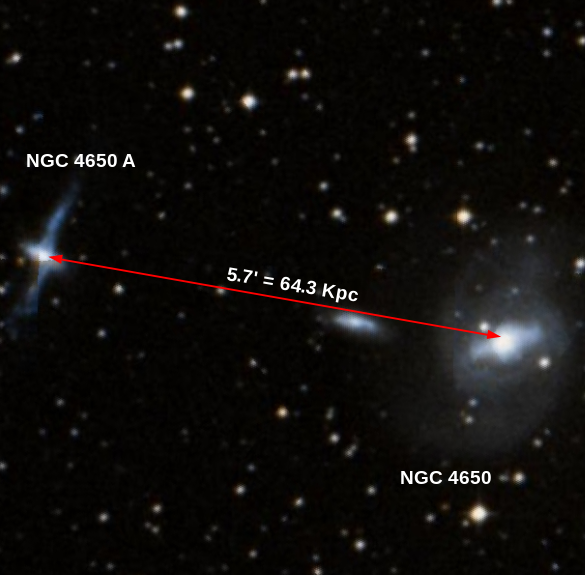}
\caption{ Projected distance between the PRG and the closest neighbor galaxy.  {\it Top panel:} WFPC2 image of the binary interacting system, NGC\,3808. The most massive cluster is pointed out with a blue circle, which is located on the forming polar ring. {\it Bottom panel:} SDSS color image for NGC\,4650A and NGC\,4650.}
\label{fig:projdist}
\end{figure}

%discussion about NGC3808
The interacting system, NGC\,3808A-B, is at an early stage in the formation of its polar ring. This is evidenced by the low surface brightness stream connecting the two galaxies (cf. Fig. \,\ref{fig:selection} and \ref{fig:projdist}). The size of the forming ring around NGC\,3808\,B is comparable to the optical diameter of the host. 
We observe the formation of young and massive compact stellar systems as a consequence of this tidal interaction evidenced by their presence not only in the main galaxy body, but also along the polar ring and the tidal bridge. These clusters are young and fairly massive with an age of $\leq1.5$\,Gyr and mass in the range $\geq10^4M_\odot - 5\times10^6M_\odot$.
The most massive clusters are candidates to become globular clusters surviving cluster dissolution processes such as galactic tidal shocking and internal stellar evolution mass loss. The tidally triggered burst of star and cluster formation is also supported by the temporal correlation of the star formation rate peaks at 1-2\,Gyr, for both galaxies, NGC\,3808\,A and the PRG NGC\,3808\,B. That is, the tidal encounter 1-2\,Gyr ago triggered a strong burst of star formation required for the formation of clusters of such mass, i.e., $\simeq130$M$_\odot$/yr (NGC\,3808\,A) and $\simeq208$M$_\odot$/yr (NGC\,3808\,B). The most massive star cluster, NGC\,3808\,B-8 (cf. Table\,\ref{tab:prop_N3808}), is spatially projected onto the tidal debris and will likely remain on a closed orbit around NGC3808\,B (cf. Fig. \ref{fig:projdist}).This will, however, have to await spectroscopic confirmation. Its apparent luminosity is $V_{F555W}=22.47$\,mag, which will make spectroscopic follow up challenging, but not impossible. NGC3808\,B-8 has a signal-to-noise of S/N$=28$, which makes it just high enough to attempt to measure its size. For that, we use the software package {\sc ishape} \citep{Larsen99}, which fits the object's profile with an analytic profile (King, EFF) using a WFPC2 PSF and a charge diffusion kernel specific to the image filter and position on the detector, which we generate with the {\sc TinyTim} software package \citep{Krist&Hook04,Krist&Hook11}. \cite{Larsen99} demonstrated that for objects with a S/N$\geq30$, one can reliably measure the object's effective radius, down to a size of 10\% of the FWHM of the stellar PSF. At the distance to NGC3808\,B and a WFPC2 FWHM$_{PSF}=1.82$\,pix this suggests that we can measure a size if the object has a intrinsic size greater than 8.54\,pc. We obtain an effective radius for NGC\,3808\,B-8 in the three filters from a best-fit King profile with a low concentration of $C=r_t/r_c=5$ of $r_{\rm eff}=18.98^{+1.71}_{-4.83}; 25.23^{+1.43}_{-2.01}; 29.50^{+2.20}_{-29.37}$\,pc in the $F450W,F555W$ and $F814W$ filters, respectively. We can therefore conclude that we are confident with the measurement in all filters, comparing with our resolution limit of 8.54\,pc. Such a mild increase ($<10\%$) of the $r_{\rm eff}$ with increasing wavelength is expected as a result of different stellar population, which have been previously observed in other compact stellar systems, such as nuclear star clusters, first by \cite{Kormendy&McClure93}, later by \cite{Matthews99} and \cite{Georgiev&Boeker14} and massive star clusters, e.g., \cite{Puzia14} and confirmed in \cite{Carson15}. NGC\,3808\,B-8 mass (${\cal M}=4.6\times10^6M_\odot$) and size ($r_{F555W, \rm eff}=25.2$\,pc) places this object in the region of observed massive and young UCDs and/or UCD progenitors. NGC3808\,B-8 has a photometric age of $t\simeq1$\,Gyr, which implies that for an assumed solar metallicity and a Kroupa IMF, the CB07 SSP models give a mass loss of about 27\% due to stellar evolution from 1 to 14\,Gyr. That is, the \emph{passive} evolution of NGC3808\,B-8 will result in a final cluster mass of $3.3\times10^6M_\odot$ as well as a decrease in size by about 30\% \citet{Baumgardt03}, i.e., $r_{F555W, \rm eff}\simeq17.5$\,pc. This will place this object in the area of observed UCDs and will support the idea that they can form during galactic tidal interactions \citep{Hilker09,Mieske12,Norris14,Taylor15}. The most massive compact object in the more massive galaxy, NGC\,3808\,A-18, has a slightly smaller mass, ${\cal M}=3.6\times10^6M_\odot$, and an older age of $t\simeq1.2$\,Gyr. We cannot measure its size because it is fainter and with a smaller S/N (cf. Table\,\ref{tab:prop_N3808}).

%Discussion about NGC4650A results
NGC\,4650\,A has a polar ring that is more than twice as extended (about 21\,Kpc in projected diameter, see Fig. \,\ref{fig:selection}) than the host galaxy  with about 7.5\,Kpc projected diameter. The polar structure in this galaxy is a disk in differential rotation and strong spiral arms \citep{arnaboldi97} whose kinematics is best reproduced when the principal plane of the dark halo is aligned with the polar disk \citep{Combes96}.  There is no indication for an ongoing interaction with a nearby companion as in the case of NGC3808\,B. The nearest companion, NGC\,4650, is nearly 65\,Kpc away (three times the diameter of the PR) and, in the next section, we discuss when a possible interaction could have taken place based on the projected distance and the epochs of peak SFR used as a lookback time. The stellar systems that we observe in NGC\,4650\,A are $\leq$5\,Gyr, and that is mostly due to incompleteness at old ages. In fact, the young clusters ($\leq$100\,Myr) and with low masses $10^3-10^4M_\odot$ are located along the polar ring and close to H\,II regions. This indicates that the most recent star formation episode in the PR proceeded at a much lower pace. Their location in the PR and low masses suggest that these objects will probably not survive for more than few ten Myrs owing to mass loss from stellar evolution and from additional tidal shocking. On the other hand, a number of clusters with ages $\geq500$\,Myr have masses larger than $5\times10^4M_\odot$ and have thus a better chance to survive as bound systems over a Hubble time and become globular clusters. The most massive clusters found in this PRG have a mass of about $5\times10^5M_\odot$ and are an order of magnitude less massive than those we observe in NGC3808\,B. This is probably largely due to the incompleteness at old ages. We thus cannot probe the earliest phase of PR formation in the same way as we witness it in NGC3808\,B. The earliest burst of star formation shown in Fig. \,\ref{fig:sfr} for NGC\,4650A (black triangles) is up to a lookback time of 5\,Gyr (the oldest cluster that we observe). The amount of HI along the polar ring is $\simeq8\times10^9M_\odot$ \citep{arnaboldi97}. This relatively high HI mass implies that star cluster and field star formation can possibly continue.

The large difference in the values of peak SFR between the two PRGs is in support of their different evolutionary stages, where NGC3808\,B is younger and in the process of formation, while NGC\,4650\,A is older. However, deeper observation in the optical and near infrared are required to obtain a more complete census of the oldest cluster populations around the PRGs and thus to probe their earliest phases of formation. It is however clear that we do not observe any classically old GCs with $M_V\lesssim-9$\,mag or brighter UCD-type objects.  

\subsection{Implications for the tidal PRG formation scenario}

Here we discuss whether the timing between the peak of the star clusters' age distribution is in accordance with a single fly-by tidal interaction or possibly with a more periodic tidal interaction between the galaxies in close binary system.

As discussed above, the ages of the star clusters in NGC\,3808\,B suggest that the massive clusters formed as consequence of a recent starburst event. We explore the possibility that the tidal interaction between NGC\,3808\,A and B led to their formation is also in general a viable mechanism that can lead to the formation of PRGs. For this analysis, we use the peak SFR age as a lookback clock for the past interaction, and the projected distance between the galaxies to calculate the travel speed that would cover that distance for that time. We compare this to numerical simulations of PRG formation. 

The current separation between the two galaxy components in the NGC\,3808 system is $\simeq27.5$\,Kpc (see Fig. \ref{fig:projdist}). Considering that the two galaxies have roughly the same radial velocities, we assume that they are moving only perpendicularly to the line of sight. Therefore, we can estimate the required tangential velocity to cover the projected distance between the two galaxies for a time that equals the most recent epoch of peak SFR (see Fig. \ref{fig:sfr}). Hence, $v_t$ is 24.5\,km/s. This velocity compares well with the expectations of the numerical simulations by \cite{Bournaud03} in which they try to explain the formation of the PRG in a tidal interaction fashion. They refer to this velocity as the relative velocity before the encounter. A simple estimation of the orbital period, using the current distance between the two galaxies as the semimajor axis and the photometric mass derived in this study, gives us a period of P\,$\simeq$\,3.5\,Gyr. 

The closest neighbor to NGC\,4650\,A, which has nearly identical radial velocity, is the S0 galaxy, NGC\,4650 (Fig. \ref{fig:projdist} bottom panel). Their projected separation is about 65\,kpc and NGC\,4650 could be a good candidate to provide a significant mass of gas for the formation of the polar ring of NGC\,4650\,A.  However, it is unclear whether this is indeed the case because earlier 21cm line observations did not observe an HI bridge between these galaxies which is expected for tidally interacting galaxies \citep{arnaboldi97,Vandriel02}. The best example for HI streams connecting interacting galaxies is the M\,81-M\,82-NGC\,3077 system, which is however much closer to us, at about 4\,Kpc, and the galaxies are more closely packed in space. Thus, the expected HI bridge connecting NGC\,4650\,A and NGC\,4650 might be stretched to low surface brightness over the large separation between the galaxies and, therefore, fall below the detection limits of those studies. Unfortunately, the ages of the star cluster population cannot probe the fly-by encounter scenario between these two galaxies because we are incomplete at old ages and thus insensitive to the earliest/first epoch of the formation of this PRG. We therefore cannot conclude in  support or against the tidal interaction scenario for NGC\,4650\,A. Deeper observations covering NIR wavelengths will greatly help to date the first burst of SF that formed the PRG and whether that is consistent with a fly-by encounter. Finding multiple peaks in the age distribution of the star cluster population, as opposed to a more continuous age distribution or a single peak, could suggest that NGC\,4650\,A may be on a closed (eccentric) orbit around NGC\,4650. With this in mind, we can calculate what might be the orbital periods assuming that both galaxies are gravitationally bound (for a\,$\simeq$\,65\,Kpc separation, the period is P\,$\simeq$\,2Gyr). If the orbital period matches the peak age of the star clusters' age distribution this would imply that it is not a single fly-by tidal interaction event, but rather episodic. However, an episodic tidal encounter between the galaxies will make the polar ring dynamically unstable and will likely be destroyed in few orbital periods. Apparently, the NGC\,4650\,A polar ring has been stable for the last Gyr or so, as evidenced by the oldest star clusters that we observed in the ring. Therefore, it is unlikely that NGC\,4650\,A is on a close orbit around NGC\,4650\,A and we favor a single fly-by event for this system.\\
For NGC3808\,A-B system, the duration of the active phase of interaction with mass transfer suggested by \cite{Reshetnikov96} is in agreement with our estimates from the analysis of the compact stellar systems. The majority of the cluster ages are within 0.2-1\,Gyr and the highest peak of star formation is attained for both NGC3808 members between 0.4\,Gyr and 1\,Gyr. According to the analysis carried out by \cite{Reshetnikov96}, these galaxies are gravitationally bound. However, with our current data (only radial velocity) we cannot confirm that the members are in closed circular orbits. An important factor that is favorable for mass transfer is the low relative velocity between the two members ($v_t=24.5\,km/s$), estimated with the information that they have nearly identical radial velocities and that our cluster-based analysis suggests a peak of SFR around 1\,Gyr. With our data alone, it is unclear, whether the mass transfer and thus its PRG in formation will remain stable.

In general we can conclude that in order to have a stable and old ($>1$\,Gyr) polar ring like that observed around NGC\,4650\,A a fly-by tidal encounter is favored over a periodic closed-orbit interaction. This could also explain why PRGs are not that commonly observed, i.e., possibly only fly-by tidal encounters with the right geometric and velocity properties of the encounter could result in the formation of a stable PR. We are not in the position to exclude the HI accretion formation scenario, however, deeper and wider-angle HI observations would show whether the PR can form from an intergalactic reservoir of HI as suggested in earlier studies \citep{Brook08,Spavone10,Iodice14}.

\section{Conclusions}\label{Sect:Conclusions}

We used archival HST/WFPC2 imaging in the optical filters $F450W, F555W$ or $F606W$ and $F814W$ to study the compact stellar systems of two galaxies, NGC\,3808B and NGC\,4650A, which are classified as classical examples of polar ring galaxies. They are in different evolutionary stages as far as the polar ring is concerned. We performed artificial star tests to determine the completeness of the data and define star cluster selection criteria based on the object profiles.  The data are used to study the color-magnitude and color-color distribution of star clusters. We have estimated the ages of the clusters based on their colors and their masses using M/L ratios from the \cite{Bruzual03} SSP models at an adopted metallicity (see \S\,\ref{subsect: SSP}). Using the cluster mass corrected for stellar evolution mass loss and the maximum cluster mass-galaxy SFR relation \citep{Weidner04,Maschberger11}, we derive the SFR and the SFH of the PRGs to test the tidal formation scenario. Our main findings are as follows:

\begin{enumerate}

\item We find young star clusters along the polar ring in NGC\,3808B with ages up to 1\,Gyr and as massive as $5\times10^6M_\odot$. We find the size of this most massive cluster, NGC\,3808\,B-8 of $r_{F555W, eff}\simeq25$\,pc. Its passive evolution placeS it in the mass-size region where UCDs are observed. In NGC\,4650\,A we observe clusters with a wide age spread (few Myr up to 5\,Gyr). The most massive star cluster in NGC\,4650\,A is about $5\times10^5M_\odot$, however, the data is severely incomplete toward older ages (cf. Fig. \ref{fig:age_mass}). We thus find evidence that the brightest most massive clusters are potential candidates to become GCs or UCDs.

\item The peak SFR derived from the most massive cluster in NGC\,3808\,B is 208M$_\odot$/yr in the age bin 1-2\,Gyr (Fig. \ref{fig:sfr}). This is confirmed by strong infrared and H$\alpha$ emission found by \cite{Reshetnikov96}. Our estimate for the SFR$_{H{_\alpha}} is \simeq 0.13M_\odot$/yr. The SFR analysis for NGC\,4650\,A shows lower peak SFRs than that in NGC\,3808\,B with a peak SFR of $10M_\odot$/yr. However, we are insensitive to the earliest periods of star formation.

\item From the analysis of the peak epoch of star formation and the apparent separation between the galaxies we conclude that a fly-by tidal encounter is possible to form a stable and old polar ring, which has been also shown by numerical models. A periodic close-orbit tidal encounter is unlikely for NGC\,4650\,A as it may well render the polar ring dynamically unstable.      
      
\end{enumerate}      

Making progress in our understanding of the formation of these PRGs, and on the formation of such galaxies in general, will require complementing the optical data with follow-up near-IR observations. This will allow us to better constrain cluster ages and masses, and assess the internal extinction and therefore break the age-metallicity degeneracy. This will enable us to properly age, date, and trace back the first interaction event that most likely lead to the PRG formation. Furthermore, new detailed spectroscopic observations are required to resolve the kinematic state of the NGC\,3808\,B PR system, as well as to confirm membership of our candidates.

\begin{acknowledgements}
Ordenes-Brice\~no would like to thank the financial support by the German Fellowship Scholarship plus for M.Sc. studies at the Universit\"at Bonn in the Argelander-Institut f\"ur Astronomie as well as CONICYT-Chile for the current PhD financial support (grant CONICYT-PCHA/Doctorado Nacional/2014-21140651). THP acknowledges the support through a FONDECYT Regular Project Grant No. 1121005 and BASAL Center for Astrophysics and Associated Technologies (PFB-06). We thank the anonymous referee for helpful suggestions.  The authors further acknowledge use of SDSS data. Funding for SDSS-III has been provided by the Alfred P. Sloan Foundation, the Participating Institutions, the National Science Foundation, and the U.S. Department of Energy Office of Science. \footnote{\href{http://www.sdss3.org/}{http://sdss3.org/}} 
We acknowledge the usage of the HyperLeda database (http://leda.univ-lyon1.fr). 
This research has made use of the NASA/IPAC Extragalactic Database (NED),  which is operated by the Jet Propulsion Laboratory, California Institute of Technology, under contract with the National Aeronautics and Space Administration.
This work made use of the {\sc Overleaf}\footnote{\href{https://www.overleaf.com}{http://overleaf.com}} platform in the preparation and collaborative writing of this paper.
\end{acknowledgements}

%-------------------------------------------------------------------

\bibliographystyle{aa}
\bibliography{tesis-ms}

\clearpage
\appendix
\input{images/N3808_tableinfo.tex}

\clearpage
\input{images/NGC4650A_tableinfo.tex}

%\section[]{Additional data tables, figures and images}
%Additional data tables, figures and images

\end{document}

%% file: table.tex
\begin{table*}
  \centering
  \begin{threeparttable}[b]
  \caption{General properties of the studied polar ring galaxies (PRGs) and their neighbor galaxy (NG).}
  \label{tab:properties}
  \renewcommand{\arraystretch}{1.3}
  \begin{tabular}{llllllllll}
  \hline
  \hline
  Property              & &   PRG    & NG \tnote{a} &    PRG    & NG \tnote{a}  \\
                          & &NGC 4650A & NGC 4650    & NGC 3808B & NGC 3808A \\
  \hline
  \\
  RA                      & [J2000.0] &12:44:49.03 & 12:44:19.63 & 11:40:44.64 & 11:40:44.21  \\
  DEC                     & [J2000.0] &-40:42:50.6 & -40:43:54.0 & +22:26:49.0 & +22:25:46.3 \\
 D\tnote{b}               & [Mpc]  & 39.8 & 39.8   & 96.8\,   & 96.8\,  \\
 m\,$-$\,M \tnote{b}        & [mag]    & 33.0\,$\pm$\,0.2 & 33.0\,$\pm$\,0.8 & 34.93\,$\pm$\,0.27 & 34.93\,$\pm$\,0.03 \\
$\upsilon_{rad}$\tnote{b} & [km/s]   & 2908.5\,$\pm$\,3.6      & 2913.3\,$\pm$\,14.3     &  7066\,$\pm$\,12    & 7076\,$\pm$\,1.3  \\
L$_{B}$                   & [L$_{\odot}$] & 2.75\,$\times$\,10$^{9}$ & 2.05\,$\times$\,10$^{10}$  & 9.04\,$\times$\,10$^{9}$  & 2.4\,$\times$\,10$^{10}$ \\
$B$ \tnote{b}          & [mag]    & 14.72\tnote{c}    & 12.60            & 15.50            & 14.45  \\
$I$ \tnote{b}          & [mag]    & 12.72\tnote{c}    & 10.41            & 13.59            & 12.93  \\
$K_{s}$ \tnote{d}      & [mag]    & 11.02             & 8.56             & 11.36            & 11.00  \\
A$_{V}$ \tnote{e}          & [mag]    & 0.312             & 0.324            & 0.072            & 0.071  \\
E$_{B\!-\!V}$ \tnote{e}      & [mag]   & 0.104             & 0.1            & 0.023            & 0.022  \\
Z           & [Z$_{\odot}$] & 0.2 \tnote{f}     &  $-$             & 0.87 \tnote{h}              & 0.75 \tnote{h}   \\
SFR$_{polar\,disk}$         & [M$_{\odot}$/yr] & 0.06 \tnote{f}    & $-$              & $-$              & $-$   \\
HI mass$_{polar\,disk}$         & [M$_{\odot}$] & 8$\times$10$^{9}$ \tnote{g}    & $-$              & $-$              & $-$   \\
Total mass PRG \tnote{h}         & [M$_{\odot}$] & 4.5$\times$10$^{10}$   & 5.6$\times$10$^{11}$             & 4.1$\times$10$^{9}$              & 9.4$\times$10$^{9}$   \\
  \hline
  \hline
  \end{tabular}
   \footnotesize{}
  \begin{tablenotes}
    \item[a] Neighbor galaxy
    \item[b] From LEDA/NED 
    \item[c] \cite{Gallagher02} 
    \item[d] From 2MASS \citep{Huchra12}
    \item[e] \cite{Schlafly11}
    \item[f] \cite{Spavone10}, for the polar disk
    \item[g] \cite{arnaboldi97}
    \item[h] Estimated in this study
  \end{tablenotes}
 \end{threeparttable}
 \end{table*}

%% file: images/N3808_tableinfo.tex
%\begin{landscape}
\begin{deluxetable}{llllllrr}
%\tabletypesize{\small}
%\rotate % For landscape mode
\setlength{\tabcolsep}{0.05in} 
\tablecolumns{8}
\tablewidth{0pc} %%% <--- This is important!!! Otherwise it wont compile!!!!
\tablecaption{\small{Photometric measurements of selected star clusters candidates for NGC 3808.}  \label{tab:prop_N3808}}
\tablehead{
\colhead{ID} &
\colhead{RA,Dec (2000)} &
\colhead{F450W$_{0}$} &
\colhead{F555W$_{0}$} &
\colhead{F814W$_{0}$} &
\colhead{V$_{0}$} &
\colhead{Age} &
\colhead{Mass} \\
\colhead{} &
\colhead{[hh:mm:ss],[dd:mm:ss]} &
\colhead{[mag]} &
\colhead{[mag]} &
\colhead{[mag]} &
\colhead{[mag]} &
\colhead{[Myr]} &
\colhead{[10$^4$M$_\odot$]} \\
\colhead{(1)} &
\colhead{(2)} &
\colhead{(3)} &
\colhead{(4)} &
\colhead{(5)} &
\colhead{(6)} &
\colhead{(7)} &
\colhead{(8)} 
}
\startdata
  NGC3808B-1 	&    11:40:43.74  22:26:44.9 	&    25.34$\pm$0.09   &   25.21$\pm$0.09 &   24.57$\pm$0.16  &   25.19$\pm$0.09    &	 240.85$\pm$48.09	&  13.98$\pm$ 1.77\\
  NGC3808B-2 	&    11:40:45.95  22:27:16.7 	&    24.87$\pm$0.11   &   24.36$\pm$0.13 &   23.82$\pm$0.12  &   24.33$\pm$0.13    &	 795.19$\pm$301.58	&  70.85$\pm$11.62\\
  NGC3808B-3 	&    11:40:44.95  22:26:42.5 	&    24.35$\pm$0.07   &   24.15$\pm$0.08 &   23.67$\pm$0.11  &   24.14$\pm$0.08    &	 249.16$\pm$75.49	&  37.49$\pm$ 6.15 \\
  NGC3808B-4 	&    11:40:45.58  22:26:59.8 	&    25.14$\pm$0.10   &   24.95$\pm$0.10 &   24.53$\pm$0.13  &   24.94$\pm$0.10    &	 187.87$\pm$76.14	&  15.84$\pm$ 2.05 \\
  NGC3808B-5 	&    11:40:45.26  22:26:47.9 	&    23.98$\pm$0.12   &   23.69$\pm$0.11 &   22.91$\pm$0.13  &   23.67$\pm$0.11    &	 583.36$\pm$329.81	& 103.59$\pm$15.54\\
  NGC3808B-6 	&    11:40:44.60  22:26:26.8 	&    25.40$\pm$0.09   &   25.07$\pm$0.11 &   24.54$\pm$0.16  &   25.05$\pm$0.11    &	 497.98$\pm$310.92	&  26.01$\pm$ 3.84 \\
  NGC3808B-7 	&    11:40:47.88  22:26:59.7 	&    24.89$\pm$0.13   &   24.80$\pm$0.09 &   24.13$\pm$0.13  &   24.79$\pm$0.09    &	  27.39$\pm$15.91	&   6.08$\pm$ 3.34\\																	 
  NGC3808B-8 	&    11:40:44.34  22:26:51.9 	&    22.98$\pm$0.09   &   22.47$\pm$0.08 &   21.60$\pm$0.08  &   22.45$\pm$0.08    &	1052.00$\pm$609.32	& 467.04$\pm$52.22\\
  NGC3808B-9 	&    11:40:43.47  22:26:35.5 	&    26.25$\pm$0.13   &   26.09$\pm$0.15 &   \nodata  	     &   26.08$\pm$0.15    &	 233.19$\pm$130.17	&   6.07$\pm$ 1.15   \\
  NGC3808B-10	&     11:40:43.95  22:26:37.9	&    25.92$\pm$0.13   &   25.79$\pm$0.15 &   \nodata  	     &   25.78$\pm$0.15    &	 211.94$\pm$113.22      &   7.76$\pm$ 2.78 \\
  NGC3808B-11	&     11:40:44.77  22:26:54.6	&    24.48$\pm$0.14   &   24.43$\pm$0.15 &   \nodata  	     &   24.42$\pm$0.15    &	  28.64$\pm$19.50       &   8.84$\pm$ 4.56 \\
  NGC3808B-12	&     11:40:45.01  22:26:40.9	&    25.42$\pm$0.11   &   25.23$\pm$0.09 &   \nodata	     &   25.22$\pm$0.09    &	 274.16$\pm$227.50      &  14.53$\pm$ 3.67 \\
  NGC3808B-13	&     11:40:45.31  22:26:48.9	&    23.68$\pm$0.17   &   23.45$\pm$0.11 &   \nodata  	     &   23.44$\pm$0.11    &	 422.02$\pm$239.78	& 103.03$\pm$17.99\\
  NGC3808B-14	&     11:40:45.39  22:26:49.4	&    24.25$\pm$0.15   &   23.72$\pm$0.13 &   \nodata  	     &   23.69$\pm$0.13    &	1316.20$\pm$737.29	& 194.16$\pm$26.26 \\
  NGC3808B-15	&     11:40:44.73  22:26:18.1	&    25.67$\pm$0.14   &   25.59$\pm$0.15 &   \nodata  	     &   25.58$\pm$0.15    &	  26.16$\pm$23.19       &   2.86$\pm$ 1.70  \\
  NGC3808B-16	&     11:40:46.72  22:26:49.9	&    25.19$\pm$0.11   &   24.78$\pm$0.12 &   \nodata	     &   24.76$\pm$0.12    &	 889.26$\pm$534.58      &  52.14$\pm$ 8.33    \\
  NGC3808B-17	&     11:40:44.48  22:26:19.0	&    25.80$\pm$0.13   &   25.55$\pm$0.11 &   \nodata	     &   25.53$\pm$0.11    &	 466.26$\pm$320.33      &  16.02$\pm$ 6.03    \\
  NGC3808A-1 	&    11:40:45.11  22:25:51.7 	&    25.39$\pm$0.13   &   25.02$\pm$0.11 &   24.57$\pm$0.13  &   25.00$\pm$0.11    &	 473.04$\pm$182.73	&  22.42$\pm$ 2.32   \\
  NGC3808A-2 	&    11:40:44.92  22:25:45.5 	&    24.45$\pm$0.10   &   24.18$\pm$0.12 &   23.45$\pm$0.09  &   24.17$\pm$0.12    &	 466.86$\pm$ 220.01	&  47.83$\pm$ 5.35   \\
  NGC3808A-3 	&   11:40:43.17  22:25:52.9  	&    22.20$\pm$0.06   &   21.99$\pm$0.05 &   21.25$\pm$0.06  &   21.97$\pm$0.05    &	 367.57$\pm$118.58	& 308.11$\pm$ 0.16 \\
  NGC3808A-4 	&   11:40:44.54  22:25:40.9  	&    22.52$\pm$0.14   &   22.32$\pm$0.12 &   21.44$\pm$0.11  &   22.31$\pm$0.12    &     390.51$\pm$284.15      & 237.34$\pm$ 0.26\\
  NGC3808A-5 	&   11:40:45.32  22:26:02.4  	&    25.22$\pm$0.14   &   24.97$\pm$0.13 &   \nodata  	     &   24.96$\pm$0.13    &	 422.32$\pm$244.52	&  21.89$\pm$ 0.27    \\
  NGC3808A-6 	&   11:40:43.75  22:25:58.5  	&    25.22$\pm$0.15   &   24.89$\pm$0.14 &   \nodata  	     &   24.87$\pm$0.14    &	 621.58$\pm$293.49	&  29.32$\pm$ 3.93    \\
  NGC3808A-7 	&   11:40:43.52  22:25:55.8  	&    25.38$\pm$0.15   &   25.10$\pm$0.16 &   \nodata  	     &   25.08$\pm$0.16    &	 491.20$\pm$315.93	&  21.16$\pm$ 3.15    \\
  NGC3808A-8 	&   11:40:45.33  22:25:41.8  	&    25.05$\pm$0.11   &   24.86$\pm$0.11 &   \nodata  	     &   24.85$\pm$0.11    &	 257.18$\pm$152.16	&  16.74$\pm$ 1.81    \\
  NGC3808A-9 	&   11:40:43.15  22:25:54.5  	&    24.79$\pm$0.13   &   24.56$\pm$0.15 &   \nodata  	     &   24.54$\pm$0.15    &	 389.65$\pm$243.38	&  30.32$\pm$ 4.21     \\
  NGC3808A-10	&    11:40:44.98  22:25:42.1 	&    24.62$\pm$0.12   &   24.46$\pm$0.13 &   \nodata  	     &   24.45$\pm$0.13    &	 226.61$\pm$212.31	&  22.94$\pm$ 2.89     \\
  NGC3808A-11	&    11:40:43.13  22:25:52.6 	&    25.39$\pm$0.14   &   25.51$\pm$0.10 &   \nodata  	     &   25.52$\pm$0.10    &	   5.22$\pm$3.23        &   0.72$\pm$ 0.0673   \\
  NGC3808A-12	&    11:40:44.49  22:25:35.6 	&    25.23$\pm$0.13   &   25.26$\pm$0.17 &   \nodata  	     &   25.27$\pm$0.17    &	   9.48$\pm$3.19        &   1.43$\pm$ 0.225   \\
  NGC3808A-13	&    11:40:42.88  22:25:45.3 	&    25.34$\pm$0.17   &   24.96$\pm$0.11 &   \nodata  	     &   24.94$\pm$0.11    &	 778.83$\pm$343.53	&  31.45$\pm$ 3.41 \\
  NGC3808A-14	&    11:40:44.35  22:25:35.9 	&    25.11$\pm$0.15   &   24.78$\pm$0.13 &   \nodata  	     &   24.76$\pm$0.13    &	 643.80$\pm$287.01	&  33.10$\pm$ 4.11 \\
  NGC3808A-15	&    11:40:44.10  22:26:03.5 	&    25.38$\pm$0.17   &   25.16$\pm$0.13 &   \nodata  	     &   25.14$\pm$0.13    &	 354.32$\pm$459.83	&  16.15$\pm$ 2.08  \\
  NGC3808A-16	&    11:40:44.64  22:25:36.7 	&    24.32$\pm$0.15   &   24.27$\pm$0.16 &   \nodata  	     &   24.27$\pm$0.16    &	  21.37$\pm$7.49        &   7.46$\pm$ 1.13  \\
  NGC3808A-17	&    11:40:44.92  22:26:10.6 	&    24.27$\pm$0.10   &   24.10$\pm$0.12 &   \nodata  	     &   24.08$\pm$0.12    &	 233.80$\pm$183.17	&  32.47$\pm$ 3.67  \\
  NGC3808A-18	&    11:40:43.65  22:25:52.8 	&    23.09$\pm$0.17   &   22.58$\pm$0.13 &   \nodata  	     &   22.55$\pm$0.13    &	1233.90$\pm$347.40	& 327.51$\pm$ 4.10 \\
  NGC3808A-19	&    11:40:43.55  22:25:49.6 	&    23.22$\pm$0.09   &   23.15$\pm$0.11 &   \nodata  	     &   23.15$\pm$0.11    &	  23.61$\pm$6.78        &  21.90$\pm$ 2.36\\
\enddata
\end{deluxetable}
%\end{landscape}

%% file: images/NGC4650A_tableinfo.tex
%\begin{landscape}
\begin{deluxetable}{llllllll}
%\tabletypesize{\scriptsize}
%\rotate % For landscape mode
\setlength{\tabcolsep}{0.05in} 
\tablecolumns{8}
\tablewidth{0pc} %%% <--- This is important!!! Otherwise it wont compile!!!!
\tablecaption{\small{Photometric measurements of selected star clusters candidates for NGC 4650A.}  \label{tab:prop_N4650A}}
\tablehead{
\colhead{ID} &
\colhead{RA,Dec (2000)} &
\colhead{F450W$_{0}$} &
\colhead{F606W$_{0}$} &
\colhead{F814W$_{0}$} &
\colhead{V$_{0}$} &
\colhead{Age} &
\colhead{Mass} \\
\colhead{} &
\colhead{[hh:mm:ss],[dd:mm:ss]} &
\colhead{[mag]} &
\colhead{[mag]} &
\colhead{[mag]} &
\colhead{[mag]} &
\colhead{[Myr]} &
\colhead{[10$^4$M$_\odot$]} \\
\colhead{(1)} &
\colhead{(2)} &
\colhead{(3)} &
\colhead{(4)} &
\colhead{(5)} &
\colhead{(6)} &
\colhead{(7)} &
\colhead{(8)} 
}
\startdata
  NGC4650A-1   & 12:44:51.21  -40:43:06.7  & 25.12$\pm$0.10 & 24.95$\pm$0.12 & 24.84$\pm$ 0.13 & 24.99$\pm$0.13 &   14.65$\pm$ 11.49 &  0.59$\pm$ 0.18 \\
  NGC4650A-2   & 12:44:50.43  -40:43:14.7  & 23.21$\pm$0.09 & 22.86$\pm$0.07 & 22.60$\pm$ 0.08 & 22.95$\pm$0.08 &  162.53$\pm$ 51.21 & 11.52$\pm$ 1.51 \\
  NGC4650A-3   & 12:44:49.23  -40:43:00.2  & 23.20$\pm$0.10 & 22.81$\pm$0.11 & 22.17$\pm$ 0.11 & 22.92$\pm$0.11 &  480.87$\pm$169.37 & 23.44$\pm$ 5.19 \\
  NGC4650A-4   & 12:44:49.27  -40:43:11.8  & 24.57$\pm$0.12 & 24.16$\pm$0.10 & 23.62$\pm$ 0.11 & 24.27$\pm$0.11 &  427.99$\pm$100.62 &  6.25$\pm$ 0.80 \\
  NGC4650A-5   & 12:44:48.84  -40:43:04.7  & 24.05$\pm$0.11 & 23.81$\pm$0.11 & 23.49$\pm$ 0.13 & 23.88$\pm$0.12 &  109.40$\pm$ 18.71 &  4.24$\pm$ 0.63 \\
  NGC4650A-6   & 12:44:49.81  -40:43:35.2  & 25.35$\pm$0.11 & 25.52$\pm$0.12 & 25.12$\pm$ 0.16 & 25.48$\pm$0.13 &    4.74$\pm$  2.15 &  0.11$\pm$ 0.09 \\
  NGC4650A-7   & 12:44:47.73  -40:43:08.1  & 24.75$\pm$0.10 & 24.03$\pm$0.09 & 23.34$\pm$ 0.11 & 24.23$\pm$0.10 & 1180.18$\pm$827.40 & 10.62$\pm$ 3.83 \\
  NGC4650A-8   & 12:44:47.25  -40:43:18.3  & 25.37$\pm$0.12 & 24.92$\pm$0.13 & 24.38$\pm$ 0.11 & 25.05$\pm$0.14 &  509.67$\pm$373.71 &  3.42$\pm$ 0.75 \\
  NGC4650A-9   & 12:44:50.69  -40:43:33.0  & 24.29$\pm$0.10 & 23.82$\pm$0.10 & 23.30$\pm$ 0.11 & 23.95$\pm$0.11 &  541.65$\pm$357.42 &  9.78$\pm$ 2.30 \\
  NGC4650A-10  & 12:44:49.59  -40:43:17.2  & 23.73$\pm$0.13 & 23.46$\pm$0.11 & 23.25$\pm$ 0.10 & 23.53$\pm$0.12 &  138.04$\pm$140.68 &  6.21$\pm$ 2.17 \\
  NGC4650A-11  & 12:44:48.65  -40:43:10.2  & 25.11$\pm$0.15 & 24.31$\pm$0.09 & 23.65$\pm$ 0.11 & 24.53$\pm$0.11 & 1214.93$\pm$906.05 &  7.62$\pm$ 1.16 \\
  NGC4650A-12  & 12:44:52.69  -40:43:30.5  & 25.44$\pm$0.13 & 25.20$\pm$0.11 &     \nodata     & 25.26$\pm$0.11 &  113.82$\pm$101.24 &  1.19$\pm$ 0.66 \\
  NGC4650A-13  & 12:44:49.64  -40:43:12.0  & 24.43$\pm$0.13 & 24.36$\pm$0.12 &     \nodata     & 24.38$\pm$0.12 &   12.70$\pm$ 10.79 &  0.80$\pm$ 1.34 \\
  NGC4650A-14  & 12:44:49.43  -40:43:12.7  & 24.25$\pm$0.09 & 24.13$\pm$0.13 &     \nodata     & 24.16$\pm$0.13 &   14.88$\pm$ 10.43 &  1.32$\pm$ 0.93 \\
  NGC4650A-15  & 12:44:48.99  -40:43:01.5  & 23.78$\pm$0.14 & 23.36$\pm$0.12 &     \nodata     & 23.47$\pm$0.12 &  436.46$\pm$252.64 & 13.19$\pm$ 2.84 \\
  NGC4650A-16  & 12:44:49.51  -40:43:21.6  & 24.32$\pm$0.13 & 23.95$\pm$0.12 &     \nodata     & 24.05$\pm$0.12 &  263.78$\pm$180.57 &  5.20$\pm$ 2.50 \\
  NGC4650A-17  & 12:44:50.06  -40:43:49.8  & 24.92$\pm$0.09 & 25.15$\pm$0.13 &     \nodata     & 25.09$\pm$0.13 &    2.77$\pm$  2.57 &  0.27$\pm$ 0.48 \\
  NGC4650A-18  & 12:44:52.22  -40:43:13.0  & 24.67$\pm$0.10 & 24.80$\pm$0.08 &     \nodata     & 24.77$\pm$0.08 &    4.04$\pm$  3.13 &  0.13$\pm$ 0.88 \\
  NGC4650A-19  & 12:44:52.09  -40:43:54.6  & 25.20$\pm$0.12 & 25.36$\pm$0.14 &     \nodata     & 25.32$\pm$0.14 &    3.91$\pm$  2.82 &  0.09$\pm$ 0.37 \\
  NGC4650A-20  & 12:44:49.07  -40:43:10.3  & 24.70$\pm$0.08 & 24.93$\pm$0.11 &     \nodata     & 24.86$\pm$0.11 &    2.78$\pm$  1.97 &  0.32$\pm$ 0.68 \\
  NGC4650A-21  & 12:44:53.35  -40:43:47.4  & 24.95$\pm$0.12 & 24.47$\pm$0.12 &     \nodata     & 24.60$\pm$0.12 &  616.95$\pm$280.93 &  5.78$\pm$ 1.05 \\
  NGC4650A-22  & 12:44:51.00  -40:42:41.7  & 24.12$\pm$0.10 & 23.30$\pm$0.10 & 22.70$\pm$ 0.11 & 23.53$\pm$0.10 & 1834.72$\pm$1264.65&  36.42$\pm$ 7.45 \\
  NGC4650A-23  & 12:44:47.08  -40:42:23.9  & 24.88$\pm$0.11 & 24.00$\pm$0.10 & 23.30$\pm$ 0.09 & 24.25$\pm$0.11 & 5105.40$\pm$3777.45&  43.79$\pm$10.66 \\
  NGC4650A-24  & 12:44:47.95  -40:42:17.7  & 23.56$\pm$0.12 & 23.28$\pm$0.12 & 22.56$\pm$ 0.11 & 23.36$\pm$0.13 &  323.28$\pm$ 207.62&  11.65$\pm$ 5.90 \\
  NGC4650A-25  & 12:44:47.41  -40:42:09.9  & 24.33$\pm$0.13 & 23.66$\pm$0.12 & 22.98$\pm$ 0.13 & 23.85$\pm$0.14 & 1152.98$\pm$880.74 &  15.75$\pm$30.36 \\
  NGC4650A-26  & 12:44:45.62  -40:42:01.3  & 23.64$\pm$0.09 & 23.44$\pm$0.07 & 23.15$\pm$ 0.11 & 23.50$\pm$0.08 &   26.60$\pm$ 15.41 &   2.87$\pm$3.24 \\
  NGC4650A-27  & 12:44:48.90  -40:42:28.7  & 24.23$\pm$0.10 & 23.53$\pm$0.11 & 23.01$\pm$ 0.11 & 23.73$\pm$0.12 & 1087.40$\pm$880.43 &  18.00$\pm$ 8.73 \\
  NGC4650A-28  & 12:44:48.73  -40:42:22.1  & 23.48$\pm$0.10 & 23.05$\pm$0.10 & 22.64$\pm$ 0.09 & 23.17$\pm$0.10 &  316.30$\pm$312.48 &  13.64$\pm$ 6.44 \\
  NGC4650A-29  & 12:44:47.33  -40:42:35.8  & 24.67$\pm$0.11 & 24.55$\pm$0.10 &     \nodata     & 24.58$\pm$0.10 &   14.84$\pm$12.52  &   0.89$\pm$ 0.95 \\
  NGC4650A-30  & 12:44:48.37  -40:42:25.5  & 23.24$\pm$0.09 & 23.14$\pm$0.11 &     \nodata     & 23.17$\pm$0.11 &   13.91$\pm$17.67  &   2.85$\pm$ 3.32 \\
  NGC4650A-31  & 12:44:45.58  -40:42:14.8  & 23.17$\pm$0.13 & 23.28$\pm$0.14 &     \nodata     & 23.25$\pm$0.14 &    3.39$\pm$2.69   &   0.56$\pm$ 3.74 \\
  NGC4650A-32  & 12:44:48.36  -40:42:39.4  & 23.30$\pm$0.14 & 22.89$\pm$0.11 &     \nodata     & 23.01$\pm$0.11 &  403.03$\pm$317.28 &  19.31$\pm$ 4.10 \\
  NGC4650A-33  & 12:44:47.00  -40:42:26.5  & 22.85$\pm$0.08 & 22.32$\pm$0.07 &     \nodata     & 22.47$\pm$0.07 &  713.95$\pm$492.60 &  44.73$\pm$ 8.27 \\
  NGC4650A-34  & 12:44:48.82  -40:42:29.3  & 23.86$\pm$0.15 & 23.50$\pm$0.12 &     \nodata     & 23.60$\pm$0.12 &  257.47$\pm$159.29 &   7.73$\pm$ 2.32 \\
  NGC4650A-35  & 12:44:45.85  -40:42:14.2  & 24.22$\pm$0.14 & 24.04$\pm$0.12 &     \nodata     & 24.09$\pm$0.12 &   53.32$\pm$30.66  &   2.49$\pm$ 1.36 \\
\enddata											                				        					   
\end{deluxetable}											        					     
%\end{landscape}